\documentclass[10pt]{article}
\usepackage[all,cmtip]{xy}
\usepackage{amsmath, amssymb,eucal,bbm}
\renewcommand{\l}{\lambda}  
\newcommand{\bR}{\mathbb{R}}  
\newcommand{\bZ}{\mathbb{Z}}

\newcommand{\ga}{\mathfrak{a}}

\newcommand{\gf}{\mathfrak{f}}  
\renewcommand{\gg}{\mathfrak{g}}

\newcommand{\gl}{\mathfrak{l}}  
  
\newcommand{\gn}{\mathfrak{n}}

\newcommand{\gu}{\mathfrak{u}}

\renewcommand{\square}{\kern1pt\vbox  
               {\hrule height 0.6pt\hbox{\vrule width 0.6pt\hskip 3pt  
    \vbox{\vskip 6pt}\hskip 3pt\vrule width 0.6pt}\hrule height0.6pt}  
    \kern1pt}  

\newcommand{\ra}{\rightarrow}

\newtheorem{Pb}{Problem}

\newtheorem{Th}{Theorem}  
\newtheorem{Prop}{Proposition}  
  
\newtheorem{Cor}{Corollary}  
\newtheorem{Lem}{Lemma}  
\newtheorem{Def}{Definition} 
\newcommand{\bP}{\begin{Pb}\ \ } 
\newcommand{\eP}{\end{Pb}}  
\newcommand{\bt}{\begin{Th}\ \ }  
\newcommand{\et}{\end{Th}}  
\newcommand{\bp}{\begin{Prop}\ \ }  
\newcommand{\ep}{\end{Prop}}  
\newcommand{\bc}{\begin{Cor}\ \ }  
\newcommand{\ec}{\end{Cor}}  
\newcommand{\bl}{\begin{Lem}\ \ }  
\newcommand{\el}{\end{Lem}}  
\newcommand{\bd}{\begin{Def}\ \ }  
\newcommand{\ed}{\end{Def}}  
  
\newcommand{\pf}{\noindent{\it Proof:\ \ }}  
\newcommand{\qed}{\hfill\square}

\newcommand{\ot}{\otimes}

\newcommand{\be}{\begin{equation}}  
\newcommand{\ee}{\end{equation}}  
  
\newcommand\re[1]{(\ref{#1})}  
\newcommand{\arr}{\begin{array}{rlll}}  
\newcommand{\ea}{\end{array}}  
\newcommand{\bea}{\begin{eqnarray}}  
\newcommand{\eea}{\end{eqnarray}}  
\newcommand{\bean}{\begin{eqnarray*}}  
\newcommand{\eean}{\end{eqnarray*}}  
  
\catcode`@=11  
\@addtoreset{equation}{section}  
\catcode`@=12

\begin{document}
%%%%%%%%%%%%%%%%%%%%%%%%%%%%%%%%%%%%%%%%%%%%%%%%%%%%%%%%%%%%%
%\begin{titlepage}
 \rightline{LTH 1003} 
 \rightline{ZMP-HH/14-4}
%\rightline{hep-th/yymmnnn}  
%\rightline{draft: \today}  
\vskip 1.5 true cm  
\begin{center}  
{\large Time-like reductions of five-dimensional supergravity}\\[.5em]
\vskip 1.0 true cm   
{V. Cort\'es$^{1}$, P. Dempster$^{2}$ and T. Mohaupt$^{2}$} \\[3pt]  
$^1${   
Department of Mathematics\\  
and Center for Mathematical Physics\\ 
University of Hamburg\\ 
Bundesstra{\ss}e 55, 
D-20146 Hamburg, Germany\\  
cortes@math.uni-hamburg.de} \\[1em]
$^2${Department of Mathematical Sciences\\ 
University of Liverpool\\
Peach Street \\
Liverpool L69 7ZL, UK\\  
pdemp@liverpool.ac.uk, Thomas.Mohaupt@liv.ac.uk
}\\[1em] 
%\today
January 22, 2014, Revised: February 19, 2014
\end{center}  
\vskip 1.0 true cm  
%%%%%%%%%%%%%%%%%%%%%%%%%%%%%%%%%%%%%%%%%%%%%%%%%%%%%%%%  
\baselineskip=18pt  
\begin{abstract}  
\noindent  

In this paper we study the scalar geometries occurring in
the dimensional reduction 
of minimal five-dimensional supergravity to three Euclidean
dimensions, and find that these depend on whether one
first reduces over space or over time. In both cases
the scalar manifold
of the reduced theory is 
described as an eight-dimensional Lie group $L$ (the Iwasawa 
subgroup of $G_{2(2)}$) with a left-invariant 
para-quaternionic-K\"ahler structure. 
We show that depending on whether one reduces
first over space or over time, the group $L$ is mapped to 
two different open $L$-orbits on the pseudo-Riemannian symmetric
space $G_{2(2)}/(SL(2) \cdot SL(2))$. These two orbits are inequivalent
in the sense that they are distinguished by the existence of
integrable $L$-invariant complex or para-complex structures.

\end{abstract}

%\end{titlepage} 

\newpage
 
\tableofcontents

%%%%%%%%%%%%%%%%%%%%%%%%%%%%%%%%%%%%%%%%%%%%%%%%%%%%%%%%%%%%

\section{Introduction}

The dimensional reduction of gravity, supergravity and string theory 
over time reveals symmetries
that are otherwise hidden, is relevant for gravitational instantons, 
and  allows one to generate stationary solutions
by subsequent dimensional lifting
\cite{Gibbons:1979xm,Breitenlohner:1987dg,Moore:1993zc,Berkooz:2008rj}. 
In the simplest examples 
the scalar manifolds of theories obtained by dimensional
reduction on tori of Lorentzian signature 
are locally symmetric Riemannian spaces with 
split signature. 
Particular cases studied in  the literature are
the symmetric spaces occurring when gravity 
coupled to matter is reduced from 
four to three dimensions  \cite{Breitenlohner:1987dg};
reductions of $D$-dimensional gravity, of bosonic and
heterotic string theory, and of eleven-dimensional supergravity 
on Lorentzian tori \cite{Hull:1998br}; and reductions
of extended four-dimensional supergravities with symmetric
target spaces over a time-like circle \cite{Gunaydin:2005mx}.

Global aspects of time-like reductions have been less studied,
but some complications have been observed in toroidal
compactifications of string theory which include a time-like
direction \cite{Moore:1993zc}. While in space-like reductions
leading to Riemannian symmetric target spaces 
$M=G/K$ of non-compact type
one can rely
on the Iwasawa decomposition 
$G=KL$, 
to provide a global parametrization of $M$ using the simply
transitive action of the solvable Iwasawa subgroup $L\subset G$,
such a global parametrization is no longer possible for the 
pseudo-Riemannian symmetric spaces $G/H$ appearing in time-like reductions. 
However examples show that it might still be 
possible to find a decomposition of 
the form $HL$ for an open subset of $U\subset G$, leading to a local
parametrization of the space $G/H$. 
In this case the Iwasawa subgroup $L$ still acts with an open orbit.
%Thus before dividing out the discrete duality symmetries, the scalar
%manifold is a proper subset of the corresponding symmetric
%space. (ii) The discrete stringy duality group $D$ acts
%ergodically, and therefore the tentative string moduli space, 
%the quotient $D \backslash G/H$, is not a Hausdorff space. 
%More recently the consequences for the classification of
%BPS and non-BPS solutions of supergravity theories have
%been investigated. 
In \cite{Bossard:2009at} it was shown 
that duality transformations relating BPS to non-BPS solutions
correspond to `singular' elements of $G$, i.e.\ elements
outside an open dense set $U\subset G$ decomposed as $U=HL$. 
In \cite{Chemissany:2010zp} it was shown that solutions with
regular event horizons correspond to complete geodesics which are 
contained within a `solv-patch', i.e.\ an open orbit of the Iwasawa 
subgroup, whereas geodesics which are not fully contained in 
a single solv-patch lift to singular space-time geometries.

In this paper we investigate further consequences of the
non-transitive action of the Iwasawa subgroup. If 
the Iwasawa subgroup acts with more than one open orbit, then
there is no a priori reason why any two open orbits should
be equivalent. 
And if open orbits are not equivalent, it becomes
necessary to decide which orbit corresponds to a given dimensional 
reduction. More specifically, we will now explain why 
this becomes an issue when 
reducing five-dimensional supergravity coupled to vector multiplets
to three Euclidean dimensions. Recall that the dimensional
reduction of four-dimensional $N=2$ vector multiplets to
three Lorentzian dimensions leads to a scalar geometry which
is quaternionic-K\"ahler \cite{Ferrara:1989ik}. The resulting
map between (projective) special K\"ahler manifolds and
quaternionic-K\"ahler manifolds is known as the $c$-map.
%If the four-dimensional theory is itself obtained
%by reduction from five dimensions one obtains a subclass
%of $c$-map spaces. 
This result extends Alekseevsky's construction \cite{MR0402649} of 
%homogeneous (not necessarily symmetric) 
symmetric and non-symmetric
quaternionic-K\"ahler
manifolds with a simply transitive solvable group of isometries
from certain K\"ahler manifolds, see also 
\cite{Cecotti:1988ad,deWit:1991nm,MR1395026}.
%with a simply transitive solvable group of isometries. 
One of the simplest examples described by Alekseevsky
is the symmetric quaternionic-K\"ahler
manifold $G_{2(2)}/SO(4)$ presented as a solvable group with
left-invariant quaternionic-K\"ahler structure. This manifold 
comprises the universal sector of five-dimensional supergravity 
reduced to three dimensions. The Alekseevsky spaces come equipped
with an integrable complex structure compatible with the 
quaternionic structure. 
More recently it was shown in \cite{Cortes:2011ut} that 
this is even true for all $c$-map spaces.

If $N=2$ vector multiplets are dimensionally
reduced with respect to time, the target space geometry is expected
to be para-quaternionic-K\"ahler instead of quaternionic-K\"ahler, as 
explained in \cite{Cortes:2003zd}. Recall that a pseudo-Riemannian
manifold $(M,g)$ of dimension $4n>4$ is called para-quaternionic-K\"ahler 
if its holonomy group is a subgroup of $Sp(\mathbbm{R}^2) \cdot
Sp(\mathbb{R}^{2n}) \subset SO(2n,2n)$ \cite{MR2140713}.
Geometrically this means that the manifold $(M,g)$ admits a parallel
subbundle $Q\subset \mbox{End}(TM)$ which is point-wise spanned by
three anti-commuting skew-symmetric endomorphisms $I,J,K=IJ$ 
such that $I^2=J^2=-K^2=\mathrm{Id}$.

In a forthcoming paper \cite{Euc4} we
prove that both the dimensional reduction of $N=2$ supergravity
with vector multiplets over time and the dimensional
reduction of Euclidean $N=2$ supergravity with vector multiplets
over space results in scalar target spaces that are 
para-quaternionic-K\"ahler. Moreover, while in the first case 
the para-quaternionic structure contains an integrable 
complex structure, it contains an integrable para-complex structure
in the second case. This indicates that when starting in five
dimensions and reducing over time and one space-like dimension,
the result will depend on the order in which the reductions are 
taken. Since this is an unexpected result, we will in this paper
investigate the simplest case, the dimensional reduction of pure
five-dimensional supergravity, in detail. We emphasize that, while
our work is motivated by \cite{Euc4}, this paper is completely
self-contained. 

The dimensional reduction of pure five-dimensional
supergravity with respect to time and one space-like dimension leads
to a scalar target space which is locally isometric to the symmetric
space 
\begin{equation}
\label{G22}
G_{2(2)}/SO_0(2,2) \simeq G_{2(2)}/ (SL(2) \cdot SL(2)),
\end{equation}
\cite{Clement:2007qy,Gaiotto:2007ag,Bouchareb:2007ax}, which
is para-quaternionic-K\"ahler. 
The classification of symmetric para-quaternionic-K\"ahler manifolds of
non-zero scalar curvature
follows from the fact that the isometry group of such a space is
simple, see Theorem 5 of \cite{MR2140713}, together with Berger's
classification of pseudo-Riemannian symmetric spaces of semi-simple
groups \cite{MR0074777,MR0074778}. The resulting list can be found
in  \cite{MR2143244,MR2681600} and contains the space
(\ref{G22}). This space represents the universal sector of 
the reduction of five-dimensional supergravity coupled to matter.
In general, the spaces obtained by such reductions
will neither be symmetric, nor even homogeneous. 
The dimensional reduction 
of five-dimensional supergravity with an arbitrary number of 
vector multiplets to 
three Euclidean dimensions will be investigated in a future
publication \cite{q-map2}.

The space (\ref{G22}) has
been studied in the literature in the context of generating
stationary solutions in four and five dimensions, in particular
stationary four-dimensional black holes \cite{Gaiotto:2007ag,Berkooz:2008rj} and
black string  solutions of five-dimensional supergravity
\cite{Compere:2009zh,Compere:2010fm}.
In  \cite{Compere:2009zh} it was verified that one obtains 
locally isometric locally symmetric spaces irrespective of whether the
reduction is carried out first over space or first over time.
It was shown in \cite{Berkooz:2008rj} that these two reductions are related to the purely space-like reduction by analytic continuation, see further comments in Section \ref{Sect_dimred}.
In this paper we will make precise the relation between 
the corresponding scalar manifolds 
and open orbits of the Iwasawa subgroup $L$  of $G_{2(2)}$ on 
$G_{2(2)}/ (SL(2) \cdot SL(2))$. We will show that while the scalar
manifolds
are locally isometric they are not related by an automorphism of $L$,
and are geometrically distinguished by the integrability properties 
of the left-invariant almost complex and para-complex structures
within the para-quaternionic structure. 

 Let us next give a more detailed summary of the results obtained
in this paper. We perform the dimensional reduction of pure five-dimensional
supergravity to three Euclidean dimensions and find that the resulting
scalar geometry 
is naturally described as a solvable Lie group $L^{(\epsilon_1, \epsilon_2)}$ 
endowed with
a left-invariant pseudo-Riemannian metric $g^{(\epsilon_1,\epsilon_2)}$ 
of split signature. The parameters $\epsilon_1, \epsilon_2 \in \{ 1, -1\}$
indicate whether the reduction 
is over a space-like
($\epsilon = -1$) or over a time-like ($\epsilon=1$) direction
in the subsequent reduction steps. 
For comparison we will also review the case of a purely space-like
reduction ($\epsilon_1=\epsilon_2=-1$). 
We find that all three groups $L^{(\epsilon_1, \epsilon_2)}$ are isomorphic
to the solvable Iwasawa subgroup of $G_{2(2)}$, 
which we will denote by $L$. 
%and therefore we will denote them simply by $L$.
In contrast to this, we prove that 
the metrics $g^{(1,-1)}$ and $g^{(-1,1)}$ are not related by an automorphism 
of the group $L$. However, we show that both pseudo-Riemannian
manifolds $(L,g^{(1,-1)})$ and $(L, g^{(-1,1)})$ can be mapped by 
a $\phi$-equivariant (respectively, $\phi'$-equivariant) isometric covering 
to open orbits 
\[ 
M=\phi(L)\cdot o, \; M'=\phi'(L) \cdot o\subset S = G/H = G_{2(2)}/(SL(2) \cdot
SL(2)),\] respectively, where $\phi, \phi':L \rightarrow G_{2(2)}$
are embeddings of $L$ into
$G_{2(2)}$ and $o=eH$ is the canonical base point of the
pseudo-Riemannian symmetric space $(S=G/H,g_S)$. 
This proves that the pseudo-Riemannian manifolds
$(L,g^{(1,-1)})$ and $(L, g^{(-1,1)})$ are locally
symmetric and locally isometric to each other. 

The symmetric space $(S,g_S)$ carries a canonical compatible $G$-invariant
para-quaternionic structure $Q$, which we will explicitly describe in 
Section \ref{SectSPQK}. This provides a direct proof that $(S,g_S,Q)$ is a 
para-quaternionic-K\"ahler manifold, as well as the open orbits
$M,M'\subset S$. Pulling back the para-quaternionic structure $Q$
by the local isometries $\phi: (L,g^{(1,-1)})\rightarrow M $, 
$\phi': (L,g^{(-1,1)})\rightarrow M' $ we obtain left-invariant
para-quaternionic-K\"ahler structures $(g^{(1,-1)}, Q^{(1,-1)})$
and $(g^{(-1,1)}, Q^{(-1,1)})$ on $L$. We show that 
$Q^{(1,-1)}$ contains a left-invariant integrable para-complex
structure $J_1=J_1^{(1,-1)}$, whereas $Q^{(-1,1)}$ contains a left-invariant
integrable complex structure $J_1=J_1^{(-1,1)}$. The structure $J_1$
is included in a standard basis $(J_1,J_2,J_3)$ of 
$Q^{(\epsilon_1,\epsilon_2)}$, which we specify explicitly on the Lie
algebra $\gl$ of $L$. 

The left-invariant structure $J_1$ is not the only left-invariant
complex ($\epsilon_1=-1$) or para-complex ($\epsilon_1=1$)
structure on $L$ which is integrable and skew-symmetric. We explicitly
describe
a second such structure $\tilde{J}_1$, commuting with $J_1$, which 
does not belong to the (para-)quaternionic structure. 

Finally we calculate the Levi-Civita
connection and curvature tensor of the metrics $g^{(\epsilon_1,\epsilon_2)}$,
in terms of a basis of left-invariant vector fields on $L$. 
Using these formulae 
we give a second proof of the fact that
the metrics $g^{(\epsilon_1,\epsilon_2)}$ are locally symmetric
and para-quaternionic-K\"ahler
by checking that the covariant derivative of the curvature tensor 
vanishes, and that $Q^{(\epsilon_1, \epsilon_2)}$ is parallel.

\section{Dimensional reduction of pure five-dimensional supergravity}\label{Sect_dimred}

%%%This is based on Paul's file Reduction of 5d supergravity

In this section we perform the dimensional reduction of pure five-dimensional
supergravity to three dimensions. The reductions over two space-like
dimensions and over one space-like and one time-like dimension will
be considered in parallel. In the latter case the time-like reduction 
can be either taken as the first or the second step. We will be interested
in comparing both options to one another. 

We start with the action for five-dimensional supergravity, coupled
to an arbitrary number $n_V^{(5)}$ of vector multiplets. In the 
conventions of \cite{Cortes:2009cs}, the bosonic part of the 
action takes the following form:
\begin{eqnarray}
S_{5}&=&\int d^{5}x\left[\sqrt{\hat{g}}\left(\frac{\hat{R}}{2}-\frac{3}{4}a_{ij}\partial_{\hat{\mu}}h^{i}\partial^{\hat{\mu}}h^{j}-\frac{1}{4}a_{ij}\mathcal{F}_{\hat{\mu}\hat{\nu}}^{i}\mathcal{F}^{j|\hat{\mu}\hat{\nu}}\right) \right. \nonumber \\
&& \left. +\frac{1}{6\sqrt{6}}c_{ijk}\varepsilon^{\hat{\mu}\hat{\nu}\hat{\rho}\hat{\sigma}\hat{\lambda}}
\mathcal{F}_{\hat{\mu}\hat{\nu}}^{i}\mathcal{F}_{\hat{\rho}\hat{\sigma}}^{j}
\mathcal{A}_{\hat{\lambda}}^{k}\right] \;.
\label{5dAction}
\end{eqnarray}
Here $\hat{\mu}, \hat{\nu}, \ldots$ are five-dimensional 
Lorentz indices and $i=0, 1,\ldots, n_V^{(5)}$ labels the five-dimensional
gauge fields. The scalars $h^i$ are understood to satisfy the constraint
\[
\mathcal{V}=c_{ijk}h^ih^jh^k = 1 \;,
\]
where $\mathcal{V}$ is a prepotential which encodes all the couplings.
While we will analyse the dimensional reduction of five-dimensional
supergravity with vector multiplets in a separate paper \cite{q-map2}, in this
article we will only consider the case of pure supergravity, where
$\mathcal{V}=(h^0)^3=1$. Then the bosonic action (\ref{5dAction}) reduces
to the one of Einstein-Maxwell theory supplemented by a Chern-Simons
term:
\begin{equation}
S = \int d^5 x \left[ \sqrt{\hat{g}}\left( \frac{\hat{R}}{2} - \frac{1}{4}
\mathcal{F}_{\hat{\mu} \hat{\nu}} \mathcal{F}^{\hat{\mu} \hat{\nu}} \right)
+ \frac{1}{6 \sqrt{6}} \varepsilon^{\hat{\mu}\hat{\nu}\hat{\rho}\hat{\sigma}
\hat{\lambda}} 
\mathcal{F}_{\hat{\mu}\hat{\nu}} \mathcal{F}_{\hat{\rho}\hat{\sigma}}
\mathcal{A}_{\hat{\lambda}} \right] \;.
\end{equation}
We perform the dimensional reduction over 2 directions by taking the metric 
ansatz $M_5=S^1\times S^1\times M_3$ with
\begin{equation}\label{MetricAnsatz}
ds_{(5)}^{2}=-\epsilon_{1}e^{2\sigma}\left(dx^{0}+\mathcal{A}^0\right)^{2}
-\epsilon_2 e^{2\phi-\sigma}\left(dx^4+B\right)^2 +e^{-2\phi-\sigma}ds^2_{(3)},
\end{equation}
where $\epsilon_{1,2}$ take the values $-1$ for reduction over a space-like 
direction and $+1$ for a time-like reduction\footnote{Note that other papers
studying reduction with respect to time, including 
\cite{Compere:2009zh,Sahay:2013xda},  use the opposite sign 
convention.}. We also introduce the 
variable $\epsilon:=-\epsilon_1\epsilon_2=(-1)^t$, where $t$ is the number of 
time-like directions in the three-dimensional theory. Note that we can take
either $x^0$ or $x^4$ to be time-like. There are two Kaluza-Klein vectors:
the four-dimensional vector ${\cal A}^0$ arising from the first reduction 
step and the three-dimensional vector $B$ arising from the second. 
It will be convenient to refer to the three different reductions
as SS-type (space-like/space-like, $\epsilon_1=\epsilon_2=-1$),
ST-type (space-like/time-like, $\epsilon_1=-1$, $\epsilon_2 =1$) and
TS-type (time-like/space-like, $\epsilon_1=1$, $\epsilon_2=-1$).

After reduction, we obtain the following three-dimensional Lagrangian:
\begin{eqnarray}\label{ExplLagr}
\mathcal{L}_3 &=& \frac{R}{2}+\frac{3}{4y^2}\epsilon_1(\partial x)^2 -\frac{3}{4y^2}(\partial y)^2 -\frac{1}{4\phi^2}(\partial\phi)^2
+\frac{1}{4\phi^2}\epsilon_1
\left(\partial\tilde{\phi}+p^I\overleftrightarrow{\partial}s_I\right)^2 \nonumber \\
& & +\frac{y^3}{12\phi}\epsilon(\partial p^0)^2 
+\frac{y}{4\phi}\epsilon_2\left(\partial p^1-x\partial p^0\right)^2 \nonumber \\
& &+\frac{3}{y^3\phi}\epsilon_2\left(\partial s_0+x\partial s_1-\frac{1}{6}x^3\partial p^0+\frac{1}{2}x^2\partial p^1\right)^2 \nonumber \\
& & +\frac{1}{y\phi}\epsilon\left(\partial s_1-\frac{1}{2}x^2\partial p^0+x\partial p^1\right)^2.
\end{eqnarray}
Here $R$ is the three-dimensional Ricci scalar which does not give rise
to local dynamics. The dynamical fields are the eight scalar fields
$x,y, \phi, \tilde{\phi}, p^0, p^1, s_0, s_1$, which have the following
five-dimensional origin: the scalars $x$ and $y$ arise by dimensional
reduction from five to four dimensions, and encode the degrees of 
freedom corresponding to the  Kaluza-Klein scalar $\sigma$ and the 
component ${\cal A}_0$ of the five-dimensional vector field ${\cal A}$.
Explicitly, we have
\[
y=6^{\frac{1}{3}}e^{\sigma}h^0, \hspace{5mm} x=2\cdot 6^{-\frac{1}{6}}\mathcal{A}_0.
\]
%Note that $x\sim {\cal A}_0$, and $y\sim h^0$, the former being
%evident from the 
%fact that $x$ is invariant under constant shifts.
Following the procedure of \cite{Cortes:2009cs}
we have absorbed the Kaluza Klein scalar $\sigma$ into %${\cal A}_0$ and 
$h^0$ to obtain scalars fitting into four-dimensional vector 
multiplets. In this formulation $x$ and $y$ are independent 
dynamical scalar fields, whereas $\sigma$ is a dependent field which
can be expressed in terms of $y$ via $e^{\sigma}=6^{-1/3}y$.
%{\bf Give explicit expressions? PD: We make no use of them, but they'll be useful for future reference elsewhere.}

%From this, and the constraint \textbf{[Ref. eqn]} we have
%\[
%e^{\sigma}=6^{\frac{1}{3}}y.
%\]
%For the remaining 5d fields, we have \textbf{[PD: Use $m,n$ or $\mu,\nu$ for 3d spacetime indices? Not used elsewhere]}
%
%Here $H_{mn}=2\partial_{[m}B_{n]}$ is the field strength associated with the second Kaluza-Klein vector.
%\\[2ex]
The scalars $\phi$ and $\tilde{\phi}$ arise from reducing the 
space-time metric from four to three dimensions. The field $\phi$ appearing in \eqref{ExplLagr} is related to the
Kaluza-Klein scalar in our ansatz (\ref{MetricAnsatz}) via $e^{2\phi}\rightarrow\phi$,
while $\tilde{\phi}$ arises from dualizing the Kaluza-Klein vector.
In particular,
\[
H_{mn}=\frac{1}{\phi^2}\epsilon_{mnp}\left(\partial^p \tilde{\phi}+p^0\overleftrightarrow{\partial}^p s_0 +p^1\overleftrightarrow{\partial}^p s_1\right),
\]
where $H_{mn}=2\partial_{[m}B_{n]}$ is the field strength associated with the second Kaluza-Klein vector.

After reduction from five to four dimensions, we have two vector fields,
namely the reduction of the five-dimensional vector field and the
Kaluza-Klein vector ${\cal A}^0$. Upon reduction to three dimensions, 
each gives rise to 2 scalars:
$p^0$ and $p^1$ correspond to the four-dimensional components
of the two vector fields,
\[
\mathcal{A}^0_{4}=-\sqrt{2} p^0, \hspace{5mm} 
\mathcal{A}_{4}=\frac{6^{\frac{1}{6}}}{\sqrt{2}}\left(p^1-xp^0\right),
\]
while $s_0$ and $s_1$ are obtained by
dualizing the vector fields after reduction to three dimensions:
\[
\mathcal{F}^0_{mn}=2\sqrt{2}B_{[m}\partial_{n]}p^0 -\epsilon\frac{6\sqrt{2}}{\phi y^3}\epsilon_{mnp}\left(\partial^p s_0 +x\partial^p s_1 +\frac{1}{2}x^2 \partial^p p^1-\frac{1}{6}x^3 \partial^p p^0\right),
\]
\begin{eqnarray*}
\mathcal{F}_{mn} &=& \frac{6^{\frac{1}{6}}}{\sqrt{2}}\left\lbrace -2B_{[m}(\partial_{n]}p^1 -x\partial_{n]}p^0)\right. \\
&&\left. \quad\; +\frac{2\epsilon_2}{\phi y}\epsilon_{mnp}\left(\partial^p s_1+x\partial^p p^1-\frac{1}{2}x^2\partial^p p^0\right)
+\sqrt{2}\mathcal{A}^0_{[m}\partial_{n]}x\right\rbrace.
\end{eqnarray*}
The scalar manifolds obtained by SS-, ST- and TS-reduction are
denoted $M^{(SS)}$, $M^{(ST)}$ and $M^{(TS)}$ respectively, and
the corresponding metrics are denoted $g^{(SS)}=g^{(-1,-1)}$, 
$g^{(ST)}=g^{(-1,1)}$ and $g^{(TS)} = g^{(1,-1)}$, respectively.

It is known that in the reduction over two space-like directions 
the eight scalars parametrize the symmetric space $G_{2(2)}/SO(4)$, 
which is quaternionic-K\"ahler. Here $G_{2(2)}$ denotes the non-compact
real form of the exceptional Lie group of type $G2$.   
It is also known that the reduction 
over one space-like and one time-like dimension gives rise 
to a space which is locally isometric to the
pseudo-Riemannian symmetric space $G_{2(2)}/(SL(2) \cdot SL(2))$,
which is para-quaternionic-K\"ahler, as expected for three-dimensional
Euclidean hypermultiplets \cite{Cortes:2003zd}. 
From (\ref{ExplLagr}) it is not manifest that reduction  
over time followed by reduction over space ($\epsilon_1=1$, $\epsilon_2=-1$)
results in the same manifold as when reducing in the opposite order
($\epsilon_1=-1$, $\epsilon_2=1$). 
%{\bf TM: Text rewritten from 
%here. Added comments on Berkooz/Pioline, adapted the discussion 
%of Compere et al.}
It is however clear that both reductions
are related to the purely space-like reduction $\epsilon_1=\epsilon_2=-1$,
and hence to one another, by analytic continuation, since
$G_{2(2)}/SO(4)$ and $G_{2(2)}/(SL(2)\cdot SL(2))$ are real forms of the
same complex-Riemannian symmetric space $G_2^{\mathbbm{C}}/SO(4,\mathbbm{C})$. 
The analytic continuations between the SS-reduction and the TS-reduction
and ST-reduction for the more general case including an arbitrary number 
of vector multiplets were given explicitly in \cite{Berkooz:2008rj}. 
Restricting to pure supergravity, and using our conventions, 
the continuation from the SS-reduction to the TS-reduction takes
the form
\begin{equation}
\label{SS2TS}
(y,x,\phi, \tilde{\phi}, p^0, p^1, s_0, s_1) \mapsto
(y,i x,\phi, i \tilde{\phi}, -i p^0,  p^1, - s_0, i s_1) \;,
\end{equation}
whilst the continuation from the SS-reduction to the ST-reduction
takes the form
\begin{equation}
\label{SS2ST}
(y,x,\phi, \tilde{\phi}, p^0, p^1, s_0, s_1) \mapsto
(y,x,\phi, -\tilde{\phi}, i p^0,  i p^1, i s_0, i s_1) \;.
\end{equation}
It is straightforward to check that these substitutions change
the relative signs of terms in (\ref{ExplLagr}) in precisely
the same way as making the corresponding changes of the parameters
$\epsilon_1$ and $\epsilon_2$. The authors of \cite{Berkooz:2008rj}
also specify a map relating the ST- and TS-reductions in their
formulae (3.16)--(3.20).  A different approach was taken 
%It is, however,
%possible to verify that one obtains the same pseudo-Riemannian symmetric
%space in both cases, at least locally. 
in \cite{Compere:2009zh}, where the 
parametrization of the scalar fields induced by dimensional reduction 
was related to a standard parametrization of the symmetric space 
$G_{2(2)}/(SL(2) \cdot SL(2))$. We will use a different parametrization which
allows us to make the (para-)quaternionic structure 
manifest, and to show that the two reductions carry additional 
geometrical structures which are not preserved by the local 
isometry relating them.

To proceed, we introduce the following basis   
for the 1-forms on the scalar manifold:
\begin{eqnarray}\label{1forms}
\eta^2 &=& \frac{1}{\phi}\left(d\tilde{\phi}+p^Ids_I-s_Idp^I\right), \hspace{5mm} \xi_2 = \frac{d\phi}{\phi},\nonumber \\
\alpha &=& \frac{\sqrt{3}}{y}dx, \hspace{5mm} \beta = \frac{\sqrt{3}}{y}dy, \nonumber \\
\eta^0 &=& \sqrt{\frac{y^3}{3\phi}}dp^0, \hspace{5mm} \eta^1 = \sqrt{\frac{y}{\phi}}\left(dp^1-x dp^0\right), \\
\xi_0 &=& 2\sqrt{\frac{3}{y^3\phi}}\left(ds_0+xds_1+\frac{1}{2}x^2 dp^1-\frac{1}{6}x^3 dp^0\right), \nonumber \\
\xi_1 &=& \frac{2}{\sqrt{y\phi}}\left(ds_1+x dp^1-\frac{1}{2}x^2 dp^0\right). \nonumber
\end{eqnarray}
These forms are also denoted %$\theta^a, a=1,\ldots, 8$. 
\begin{equation}
\label{DVbasis}
(\theta^a ) = (\eta^2,\xi_2,\alpha, \beta,\eta^0,\eta^1,\xi_0,\xi_1) \;.
\end{equation}

The metric $g$ on the target manifold associated with the 
Lagrangian \eqref{ExplLagr} then takes the form 
\begin{equation}
\label{Metric1}
4 g = 
-\epsilon_1\eta^2\otimes\eta^2
+\xi_2\otimes\xi_2 
- \epsilon_1 \alpha\otimes\alpha 
+\beta\otimes\beta
 -\epsilon \eta^0\otimes \eta^0
-\epsilon_2\eta^1\otimes\eta^1 
- \epsilon_2\xi_0\otimes\xi_0 
-\epsilon \xi_1\otimes\xi_1.
\end{equation}
%{\bf TM: I have re-arranged terms in the above formula according
%to the ordering of the one-forms introduced above. I have also
%added the following remark on analytic continuation.}
%{\bf I reversed the overall sign compared to Paul's file, so that $g$
%is positive definite for space-like reduction}. \\[2ex]
Note that under the analytic continuations (\ref{SS2TS}) and
(\ref{SS2ST}) the one-forms (\ref{DVbasis}) transform as
\[
(\eta^2, \xi_2, \alpha, \beta, \eta^0, \eta^1, \xi_0, \xi_1)
\mapsto 
(i\eta^2, \xi_2, i \alpha, \beta, -i \eta^0, \eta^1, -\xi_0, i\xi_1),
\]
\[
(\eta^2, \xi_2, \alpha, \beta, \eta^0, \eta^1, \xi_0, \xi_1)
\mapsto 
(-\eta^2, \xi_2, \alpha, \beta, i\eta^0, i\eta^1, i\xi_0, i\xi_1),
\]
which flips the relative signs in (\ref{Metric1}) in the same
way as making the corresponding changes in the parameters $\epsilon_1$
and $\epsilon_2$. 

The one-forms $\theta^a$ have the following exterior
derivatives:
\begin{eqnarray}\label{ExteriorAlg}
d\eta^2 &=& -\xi_0\wedge\eta^0 -\xi_1\wedge\eta^1 -\xi_2\wedge\eta^2, \nonumber \\
d\xi_2 &=& 0, \nonumber \\
d\alpha &=& \frac{1}{\sqrt{3}}\alpha\wedge\beta, \nonumber \\
d\beta &=& 0, \nonumber \\
d\eta^0 &=& \frac{\sqrt{3}}{2}\beta\wedge\eta^0 -\frac{1}{2}\xi_2\wedge\eta^0, \\
d\eta^1 &=& \frac{1}{2\sqrt{3}}\beta\wedge\eta^1 -\frac{1}{2}\xi_2\wedge\eta^1 -\alpha\wedge\eta^0, \nonumber \\
d\xi_0 &=& -\frac{\sqrt{3}}{2}\beta\wedge\xi_0 -\frac{1}{2}\xi_2\wedge\xi_0 +\alpha\wedge\xi_1, \nonumber \\
d\xi_1 &=& -\frac{1}{2\sqrt{3}}\beta\wedge\xi_1 -\frac{1}{2}\xi_2\wedge\xi_1 +\frac{2}{\sqrt{3}}\alpha\wedge\eta^1. \nonumber
\end{eqnarray}
This shows that they form a Lie algebra and
that $g$ can be considered as a left-invariant pseudo-Riemannian metric on
the corresponding simply connected Lie group, which is parametrized by
$(x,y,\phi,\tilde{\phi},p^0,p^1,s_0,s_1)$.
The structure constants of this Lie algebra can be read off from
the relation $d\theta^c=-c^c_{ab}\theta^a\wedge \theta^b$. 
The relations for the dual vector fields $T_a$, 
where $\langle \theta^a, T_b \rangle = \delta^a_b$, 
which we identify
with the Lie algebra generators, are 
$[T_a, T_b] = c^c_{ab} T_c$.
 
Denoting the basis dual to $(\theta^a)$ by 
\begin{equation}
\label{Vbasis}
( T_a ) = (V_2, U^2,A,B,V_0,V_1,U^0,U^1),
\end{equation}
%{\bf Ordering of the basis changed relative to Paul's file by swapping
%$A$ and $B$, to be consistent with Vicente's ordering}\\[2ex]
we obtain:
\begin{eqnarray}
\left[B, A\right]&=&\frac{1}{\sqrt{3}}A, \hspace{5mm} \left[U^2 ,V_2\right]=V_2, \nonumber \\
\left[V_0, U^0\right]&=&-V_2, \hspace{5mm} \left[V_1, U^1\right]=-V_2, \nonumber \\
\left[U^2, V_I\right]&=&\frac{1}{2}V_I\quad\textrm{for}\;I=0,1, \hspace{5mm} \left[U^2,U^I\right]=\frac{1}{2}U^I\quad\textrm{for}\;I=0,1, \nonumber \\
\left[B, V_0\right]&=&-\frac{\sqrt{3}}{2}V_0, \hspace{3mm} \left[B, V_1\right]=-\frac{1}{2\sqrt{3}}V_1, \hspace{3mm} \left[B, U^0\right]=\frac{\sqrt{3}}{2}U^0, \hspace{3mm} \nonumber \\
 & & \left[B, U^1\right]=\frac{1}{2\sqrt{3}}U^1, \nonumber \\
\left[A, V_0\right]&=& V_1, \hspace{5mm} 
\left[A, U^1\right]=-U^0, \hspace{5mm}
\left[A, V_1\right]=-\frac{2}{\sqrt{3}}U^1.
\label{Iwasawa1}
\end{eqnarray}
This Lie algebra is easily seen to be a solvable Lie algebra. 
As we will see below, it is an Iwasawa
subalgebra of the Lie algebra of $G_{2(2)}$. Thus the three dimensional
reductions provide us with scalar manifolds which can all be identified
with the group manifold $L$ of an 
Iwasawa subgroup of $G_{2(2)}$. For each of the three reductions this
manifold is equipped with a different left-invariant metric. 
The signature is, using the ordering (\ref{DVbasis}),
\begin{equation}
\label{signature}
\mbox{sign}(g) = (-\epsilon_1, +, -\epsilon_1,+, -\epsilon, -\epsilon_2,
-\epsilon_2, -\epsilon) \;.
\end{equation}
Thus for an SS reduction the metric is positive definite, while
for ST and TS reductions we obtain  split  (i.e.\ neutral) signature metrics,
but with a different distribution of $(+)$-signs and $(-)$-signs. 
Note that while scalar products are classified up to isomorphism 
by their signatures, this does not imply the existence of an 
isometry which simultaneously preserves the Lie algebra structure.
This will be important in the following. 

%\section{Geometric structures on the Iwasawa subgroup of 
%the noncompact group of type G2}

%This is taken from Vicentes file

\section{The group $G_{2(2)}$, its Iwasawa subgroup, and the symmetric space $S=G_{2(2)}/(SL(2)\cdot SL(2))$}

\subsection{The noncompact group of type G2} 
Let us denote by $G=G_{2(2)}$ the simply connected noncompact form of the 
simple Lie group of type G2. Its Lie algebra $\gg$ can be described
as follows, see \cite{MR1349140}, Ch.\ 5, Section 1.2. 
It contains $\mathfrak{sl}(V)$ as a subalgebra, where $V=\bR^3$, 
such that under the adjoint representation of $\mathfrak{sl}(V)$ 
on $\gg$ we have the following decomposition  
\[ \gg = V + \mathfrak{sl}(V) + V^*,\] 
as a direct sum of irreducible $\mathfrak{sl}(V)$-submodules. 
The remaining Lie brackets are given by
\begin{eqnarray*} [x,y] &=& -2 x\times y ,\\
{[}\xi , \eta {]} &=& 2 \xi \times \eta ,\\
{[}x ,\xi {]} &=& 3x\ot \xi -\xi (x)\mathrm{Id}\in  \mathfrak{sl}(V)\subset 
\mathfrak{gl}(V)\cong V\ot V^*, 
\end{eqnarray*} 
for all $x, y\in V$, $\xi ,\eta \in V^*$.  The  
cross products are defined by
\[ x\times y = \det (x,y,\cdot )\in V^*,\quad  \xi\times \eta = 
{\det}^{-1}(\xi ,
\eta , \cdot )\in V^{**}=V,\]
where ${\det}^{-1}\in \wedge^3V$ is the inverse of $\det \in \wedge^3V^*$.
Let us denote by $\ga$ the Cartan subalgebra of 
$\gg$ which consists of all diagonal matrices in $\mathfrak{sl}(V)$. 
We shall denote by $(e_i) = (e_1,e_2,e_3)$ the standard basis
of $V$, by $(e^i)$ its dual basis and by $e_i^j$ the endomorphism
$e_i\ot e^j$ of $V$. With this notation, 
\[ \ga = \{ \sum \l_ie_i^i| \sum \l_i=0\}.\]

\subsection{The symmetric para-quaternionic-K\"ahler manifold \\ $S=G_{2(2)}/(SL_2\cdot SL_2)$ \label{SectSPQK}} 
\bp The Lie algebra $\gg$ admits the following
$\bZ_2$-grading 
\be \label{Z2gradingEqu} \gg = \gg_{ev} + \gg_{odd},\ee
where 
\begin{eqnarray*} 
\gg_{ev} &=& \ga + \mathrm{span} \{ e_3,e^3,e_1^2,e_2^1\} \cong \mathfrak{sl}_2 
\oplus \mathfrak{sl}_2,\\
\gg_{odd} &=& \mathrm{span} \{ e_1,e_2,e^1,e^2,e_1^3,e_2^3,e_3^1,e_3^2\} .
\end{eqnarray*}
The corresponding symmetric space 
$S=G/G_{ev}$ admits a $G$-invariant 
para-quaternionic-K\"ahler structure $(g,Q)$, unique up to scale. 
The metric $g$ is induced by a multiple of the Killing form. 
\ep    

\pf It is straightforward to check that \re{Z2gradingEqu} is a
$\bZ_2$-grading of the Lie algebra $\gg$. This shows that 
$S$ is a symmetric space. 
Furthermore,  
\[ (\mathbf{h}=[e_1^2,e_2^1]=e_1^1-e_2^2,\; \mathbf{e}=e_1^2,\; 
\mathbf{f}=e_2^1),\] 
is an $\mathfrak{sl}_2$-triple $(\mathbf{h}, \mathbf{e}, \mathbf{f})$,
as well as 
\[ ([e_3,e^3]= -e_1^1-e_2^2+2e_3^3,\; e_3,\; e^3).\] 
They generate two
complementary ideals $\mathfrak{sl}_2^{(a)}\cong \mathfrak{sl}_2$, $a=1,2$, 
in $\gg_{ev}$. One can further check that the 
isotropy representation of $S$ is a tensor product
$\bR^2\ot \bR^4$ of irreducible representations of the two $SL_2$-factors.
The total irreducibility of the isotropy representation implies that the 
metric induced by the Killing form is the only $G$-invariant 
pseudo-Riemannian metric $g$ on $S$, up to scale. 
The isotropy representation of $\mathfrak{sl}_2^{(1)}\subset \gg_{ev}$ 
on the first factor 
defines a $G$-invariant almost para-quaternionic structure $Q$ on $S$, 
which consists of skew-symmetric endomorphisms. 
It is the only $G$-invariant almost para-quaternionic structure on $S$, since
the $G_{ev}$-invariant decomposition $T_oS \cong \bR^2\ot \bR^4$ is unique, where
$o=eG_{ev}$ stands for the canonical base point of $S=G/G_{ev}$.   
Moreover, $Q$ is invariant under 
parallel transport because the isotropy group coincides with the 
holonomy group of the simply connected pseudo-Riemannian symmetric 
space $S$, as a consequence of the Ambrose-Singer theorem. So 
$(S,g,Q)$ is a para-quaternionic-K\"ahler manifold. 
\qed

\subsection{The solvable Iwasawa subgroup $L\subset G$}
Let us define  
\[ \gn := \mathrm{span}\{ e_1,e^2,e^3,e_1^2,e_1^3,e_2^3\}\subset \gg .\]
We claim that $\gn \subset \gg$ is a maximal unipotent\footnote{A subalgebra
of a linear Lie algebra is called unipotent if it operates on the given 
vector space by upper triangular matrices with vanishing diagonal elements.
Note that a nilpotent Lie algebra is not
automatically unipotent if the Lie algebra is represented by matrices.
For the adjoint representation it is true that nilpotent Lie algebras
are realized as unipotent linear Lie algebras, but this is not necessarily true
for other representations. Since the representation we use is not
the adjoint representation of $\mathfrak{n}$, but the restriction 
of the adjoint representation of $\mathfrak{g}$ to $\mathfrak{n}$, 
the distinction between nilpotent and 
unipotent subalgebras is relevant.}
subalgebra
normalized by the Cartan subalgebra $\ga\subset \gg$. In fact, $\gn$ is 
precisely the
sum of the positive root spaces of $\ga$ with respect to the 
Weyl chamber containing the element $3e_1^1-e_2^2-2e_3^3\in \ga$. 
As a consequence, we obtain: 
\bp The solvable Lie algebra 
\[ \gl = \ga + \gn \subset \gg\]
is a maximal triangular subalgebra of $\gg$.
\ep 
Any maximal triangular subalgebra of $\gg$ will 
be called an \emph{Iwasawa subalgebra}, since it is the solvable
Lie algebra appearing in the Iwasawa decomposition of $\gg$. 
Any two Iwasawa subalgebras of $\gg$ are conjugated. 

The Iwasawa decomposition implies that 
the Lie subgroup $L\subset G$ with the Lie algebra $\gl \subset \gg$ 
acts simply transitively on the quaternionic-K\"ahler 
symmetric space $G/SO_4$. Therefore, the quaternionic-K\"ahler structure can be described as a left-invariant structure
on $L$. This was done in \cite{MR0402649}. Correcting some 
misprints and changing slightly the notation, 
the Lie algebra of the simply transitive group 
described by Alekseevsky is spanned by a basis %\vspace{-3mm}
\[ (G_0,H_0,G_1,H_1,\tilde{P}_-,\tilde{P}_+,\tilde{Q}_-,\tilde{Q}_+),\]%\vspace{-3mm}
with the following nontrivial brackets:
\begin{align*}
 &{[}H_0,G_0{]} = G_0,\; {[}H_1,G_1{]} = \frac{1}{\sqrt{3}}G_1,
\\ 
&{[}H_0,\tilde{U}{]} = \frac{1}{2}\tilde{U},\quad \mbox{for all}\quad 
\tilde{U} \in
\tilde{\mathfrak{u}} := \mathrm{span}\{ 
\tilde{P}_-,\tilde{P}_+,\tilde{Q}_-,\tilde{Q}_+\},\\
& {[}H_1,\tilde{P}_-{]} = \frac{\sqrt{3}}{2}\tilde{P}_- ,  
{[}H_1,\tilde{P}_+{]} = \frac{1}{2\sqrt{3}}\tilde{P}_+, 
{[}H_1,\tilde{Q}_-{]} = \!-\frac{\sqrt{3}}{2}\tilde{Q}_-, 
{[}H_1,\tilde{Q}_+{]} =\!-\frac{1}{2\sqrt{3}}\tilde{Q}_+,\\
&  {[}G_1,\tilde{P}_+{]} = -\tilde{P}_-,\; 
{[}G_1,\tilde{Q}_-{]} = \tilde{Q}_+,\; 
{[}G_1,\tilde{Q}_+{]} =\frac{2}{\sqrt{3}}\tilde{P}_+,\\
& {[}\tilde{Q}_-,\tilde{P}_-{]} = {[}\tilde{Q}_+,\tilde{P}_+{]} =G_0. 
\end{align*}
\bp The Lie algebra $\gl$ admits a basis\\ $(G_0,H_0,G_1,H_1,\tilde{P}_-,\tilde{P}_+,\tilde{Q}_-,\tilde{Q}_+)$ with the above 
commutators. 
\ep 

\pf It suffices to define
\begin{eqnarray*}
&&G_0 := -3 e_1^3,\; H_0 := \frac{1}{2}(e_1^1-e_3^3),\;  G_1:= \frac{1}{\sqrt{3}}e^2,\;H_1 :=
\frac{1}{2\sqrt{3}}(e_1^1-2e_2^2+e_3^3)\\
&&\tilde{P}_- := \sqrt{3} e_1^2,\; \tilde{P}_+ := e_1,\; \tilde{Q}_- := 
\sqrt{3} e_2^3,\; \tilde{Q}_+ := e^3.
\end{eqnarray*}%\vspace{-4mm}
\qed
    
To compare with the results obtained by dimensional reduction it is more 
convenient to work with the following basis: 
\[ ({\cal V}_1 , \ldots, {\cal V}_8 ) =
(G_0,H_0,G_1,H_1,\tilde{Q}_-,\tilde{Q}_+,-\tilde{P}_-,-\tilde{P}_+,),\]
i.e.%\vspace{-4mm}
\begin{eqnarray} %\label{Vbasis1}
&&{\cal V}_1 = -3 e_1^3,\; {\cal V}_2 = \frac{1}{2}(e_1^1-e_3^3),\;  
{\cal V}_3= 
\frac{1}{\sqrt{3}}e^2,\;
{\cal V}_4 =
\frac{1}{2\sqrt{3}}(e_1^1-2e_2^2+e_3^3),  \nonumber \\
&&{\cal V}_5 = 
\sqrt{3} e_2^3,\; 
{\cal V}_6 = e^3,\; 
{\cal V}_7:= -\sqrt{3} e_1^2,\; 
{\cal V}_8 := -e_1, \label{V-basis}
%&&V_2 = -3 e_1^3,\; U^2 = \frac{1}{2}(e_1^1-e_3^3),\;  A= 
%\frac{1}{\sqrt{3}}e^2,\;B =
%\frac{1}{2\sqrt{3}}(e_1^1-2e_2^2+e_3^3)\\
%&&V_0 = 
%\sqrt{3} e_2^3,\; V_1 = e^3,\; U^0 := -\sqrt{3} e_1^2,\; U^1 := -e_1,
\end{eqnarray}
which has precisely the same nontrivial brackets (\ref{Iwasawa1}) 
as the basis $T_a$ of the Lie algebra obtained
from dimensional reduction.
%We will see later that while the basis
%$\{ T_a \}$ obtained by dimensional reduction is
%orthonormal up to scale, the basis $\{ {\cal V}_a \}$ 
%is not orthogonal with respect to metric of the symmetric space
%the Killing form. However
%both bases can be related, 
%for all three cases of dimensional reduction, by inner 
%automorphisms of $\mathfrak{l}$. 

%\section{Two Iwasawa subgroups with open orbit and their relation
%to dimensional reduction}

\section{Realization of the scalar manifolds of the reduced theories as open orbits in the symmetric space $S$}\label{SectRealization}

Our goal is to realize the scalar manifolds $M^{(TS)}$ and $M^{(ST)}$
of the reduced theories as open orbits $M_1 = L'\cdot o$ and 
$M_2 = L'' \cdot o$ 
of Iwasawa subgroups $L',L''\subset G$ on the symmetric space $S=G/G_{ev}$.
Notice that the standard Iwasawa subgroup $L\subset G$ acts transitively
on the Riemannian symmetric space $G/SO_4$, but that the orbit $L\cdot o$ of
the canonical base point $o\in G/(SL(2)\cdot SL(2))$ under this group
is not even open. Our strategy is to look for a conjugate subgroup
$L'= C_a(L) = a L a^{-1}$, $a\in G$, such that the orbit $M_1=L' \cdot o$ is open, and 
then to try to 
show that $M_1$ is isometrically covered, up to a positive scale factor,
by at least one of the two scalar 
manifolds $M^{(TS)}$ or $M^{(ST)}$. 

In the following subsection we construct an Iwasawa subgroup $L'=C_a(L)
\subset G$
for which the orbit $L'\cdot o \subset S$ is open. Composing the isomorphism
$C_a: L \rightarrow L'$ defined by conjugation by $a\in G$ with the 
covering $L'\rightarrow M_1$ given by the 
orbit map $x \mapsto x\cdot o$ we obtain a 
$C_a$-equivariant covering
$\phi_1: L \rightarrow M_1$ and a left-invariant metric 
$g_1= \phi_1^* g_S$ on $L$, which we can compare to the metrics 
$g^{(TS)}=g^{(1,-1)}$ and $g^{(ST)}=g^{(-1,1)}$. Recall that the
solvable Lie algebra obtained from dimensional reduction
comes equipped with the basis ${\cal T}:= 
(T_1, \ldots, T_8)$, whereas the standard
Iwasawa Lie subalgebra $\gl \subset \gg$ is equipped with the
basis ${\cal V}:= ({\cal V}_1, \ldots, {\cal V}_8)$. 
Since these bases have the same structure
constants, we can identify the two Lie algebras.  
In Proposition \ref{PropGram} we compute the Gram matrix ${\cal G}_1$
of $g_1$
with respect to the basis ${\cal V}$. Contrary to the metrics
$g^{(TS)}$ and $g^{(ST)}$, for which the basis ${\cal T}$ is 
orthonormal (up to an overall factor), we find that ${\cal G}_1$
is not even diagonal. Notice that nevertheless the left-invariant
metrics $g_1$ and $g^{(TS)}$ or $g^{(ST)}$ on $L$ could be equivalent, 
that is related by an automorphism of $L$, up to a positive
scale factor. In Subsection \ref{Auto}
we determine the group $\mbox{Aut}(L)$  of all automorphisms of $L$. 
As a result 
we find in particular that for the connected component of the identity
$\mbox{Aut}_0(L) = \mbox{Inn}(L) \cong L$. In Subsection \ref{IdTS}
we prove that the metrics $g_1$ and $g^{(TS)}$ are equivalent and, more
precisely, related by a unique inner automorphism of $L$, and multiplication
by a factor of 2. Similarly 
in Subsection \ref{IdST} we construct a second Iwasawa subgroup 
$L''\subset G$ such that $M_2 = L''\cdot o$ is open and a 
covering $\phi_2: L \rightarrow M_2$ which is equivariant with respect
to an isomorphism $L \rightarrow L''$. Finally, the left-invariant 
metric $g_2 = \phi_2^* g_S$ is shown to be related to the metric $g^{(ST)}$
by a unique inner automorphism of $L$, and multiplication by a factor of 2. 
We also show that, surprisingly, 
the metrics $g_1$ and $g_2$ are not related by any automorphism. 
We will see in Section \ref{GeomStruct} that the metric Lie groups $(L,g_1)$ 
and $(L,g_2)$ 
have different geometric properties.

\subsection{Iwasawa subgroups of $G$ with an open orbit on $S$}
\label{IwaSub}

Given a subgroup $U\subset G$ we can consider the 
orbit $U\cdot o\subset S=G/G_{ev}$ 
of the canonical base point $o\in S$. 
The orbit is open if and only if $\gg_{ev} + \mbox{Lie}(U) = \gg$. 
For an Iwasawa subalgebra $\gl' = \mathrm{Ad}_a\gl\subset \gg$, $a\in G$, 
this is  the case if and only if $\gg_{ev}\cap \gl'=0$.  In that 
case, the orbit map $L' \ra M=L'\cdot o\subset S$ is a covering and we obtain 
a left-invariant locally symmetric para-quaternionic-K\"ahler structure
on $L'\cong L$ induced from the symmetric para-quaternionic-K\"ahler structure
on $S$. Notice that the orbit $L\cdot o$ (the case $a=e$) is not 
open, since $\gg_{ev}\cap \gl\neq 0$. 
\bp The element $a=\exp \xi$, where $\xi = e^1+e_3^1\in \gg$, defines
an Iwasawa subalgebra $\gl' = \mathrm{Ad}_a\gl\subset \gg$  
transversal to $\gg_{ev}$. 
\ep 
\pf We first compute $X' := \mathrm{Ad}_aX=e^{ad_\xi}X$ for every element 
$X\in \gl$. For $H=\sum \l_ie_i^i\in \ga\subset \gl$, 
\[ ad_\xi H = -[H,\xi ] = -(-\l_1e^1+ (-\l_1 +\l_3)e_3^1)= \l_1e^1 + 
(\l_1-\l_3)e_3^1,\;  ad_\xi^2H = 0,\]
implies 
\be \label{1stprimeEqu} H' = H + \l_1e^1 + 
(\l_1-\l_3)e_3^1.\ee 
Next, 
\begin{eqnarray*} ad_\xi e_1 &=& -2e_1^1+e_2^2+e_3^3 + e_3,\\ 
ad_\xi^2e_1 &=& 
-2[e^1, e_1^1]+[e^1,e_3]  -2[e_3^1,e_1^1] +[e_3^1,e_3^3] \\
&=&
-2e^1-3e_3^1-2e_3^1-e_3^1= -2(e^1+3e_3^1),\\
ad_\xi^3e_1 &=&0,
\end{eqnarray*}
implies 
\be e_1' = e_1 -2e_1^1+e_2^2+e_3^3 + e_3- e^1-3e_3^1 .\ee 
\[ ad_\xi e^2 = 2e_3,\; ad_\xi^2e^2=-6e_3^1,\; ad_\xi^3e^2=0 \implies\] 
\be {e^2}' = e^2 + 2e_3 -3e_3^1.\ee 
\[ ad_\xi e^3 = -2e_2-e^1,\; ad_\xi^2e^3 = 6 e_2^1,\; ad_\xi^3e^3=0 \implies \]
\be {e^3}' = e^3-2e_2-e^1+3e_2^1.\ee
\[ ad_\xi e_1^2 = e^2+e_3^2,\; ad_\xi^2e_1^2=2e_3,\; ad_\xi^3e_1^2=-6e_3^1,\; 
ad_\xi^4e_1^2=0 \implies\] 
\be {e_1^2}'= e_1^2 +e^2 +e_3^2+e_3-e_3^1.\ee
\[ ad_\xi e_1^3 = e^3+e_3^3-e_1^1,\; ad_\xi^2e_1^3=-2(e_2+e^1+e_3^1),\; ad_\xi^3e_1^3
=6e_2^1,\; 
ad_\xi^4e_1^3=0 \implies\] 
\be {e_1^3}' = e_1^3+e^3+e_3^3-e_1^1-e_2-e^1-e_3^1+e_2^1.\ee
\[ ad_\xi e_2^3 = -e_2^1,\; ad_\xi^2e_1^3=0\implies\]
\be \label{lastprimeEqu} {e_2^3}' = e_2^3-e_2^1.\ee 

Next we check the transversality of $\gl'$. 
Let us denote by $\pi : \gg \ra \gg_{odd}$ the projection along
$\gg_{ev}$ and by $\varphi : \gl \ra \gg_{odd}$ the map
$X\mapsto \pi (X')$.  {} From \re{1stprimeEqu}-\re{lastprimeEqu} we can read 
off $\varphi$: 
\begin{eqnarray} &&\varphi (H) = \l_1e^1+(\l_1-\l_3)e_3^1,\quad \mbox{for all}
\quad
H=\sum \l_ie_i^i\in \ga ,\nonumber \\ 
&&\varphi (e_1) = e_1 -e^1-3e_3^1,\; 
\varphi (e^2) = e^2 -3e_3^1,\; \varphi (e^3) = -2e_2-e^1, \nonumber\\
&&\varphi (e_1^2)=
e^2+e_3^2-e_3^1,\; \varphi (e_1^3) = e_1^3-e_2-e^1-e_3^1,\; 
\varphi (e_2^3)= e_2^3, \label{varphiEqu}
\end{eqnarray}
which shows that $\varphi : \gl \ra \gg_{odd}$ is an isomorphism
of vector spaces. This implies that $\gl'$ is transversal to
$\gg_{ev}$.   
\qed

Next we compute the left-invariant metric $g_1$ 
on $L\cong L'$ which corresponds to
the locally symmetric para-quaternionic-K\"ahler manifold 
$M_1=L'\cdot o\subset S$. Let us denote by $B$ the Killing form
of $\gg$ and by $\langle \cdot ,\cdot \rangle_B$
the scalar product on $\gg_{odd}$ obtained by restricting 
$\frac{1}{8}B$. 
\bl \label{KillingLemma} 
The nontrivial scalar products between elements of 
the basis\linebreak 
$(e_1,e_2,e^1,e^2,e_1^3,e_2^3,e_3^1,e_3^2)$ of 
$\gg_{odd}$ are precisely:
\[ \langle e^1 , e_1 \rangle_B =\langle e^2 , e_2 \rangle_B = 3,\;
\langle e_2^3 , e_3^2 \rangle_B =\langle e_1^3 , e_3^1 \rangle_B =1.
\]
\el 

The scalar product 
$\langle\cdot ,\cdot \rangle_1$ on $\gl$  which defines the 
metric $g_1$ is precisely the pull back of  $\langle \cdot ,\cdot \rangle_B$
by the isomorphism $\varphi = \pi \circ \mathrm{Ad}_a  : \gl \ra \gg_{odd}$.  
\bp \label{PropGram}
The matrix representing the scalar product $\langle\cdot ,\cdot \rangle_1
= \varphi^* \langle \cdot ,\cdot \rangle_B$ in the basis ${\cal V}$ is:
\begin{equation}
\label{Gramlprime}
%G(\mathfrak{l}') 
{\cal G}_1= \left( \begin{array}{rrrrrrrr} -18&-3&6\sqrt{3}&0&0&0&-12\sqrt{3}&-18\\
-3&0&0&0&0&0&0&-\frac{3}{2}\\
6\sqrt{3}&0&0&0&0&-2\sqrt{3}&0&0\\
0&0&0&0&0&0&0&-\frac{\sqrt{3}}{2}\\
0&0&0&0&0&0&-3&0\\
0&0&-2\sqrt{3}&0&0&0&6\sqrt{3}&3\\
-12\sqrt{3}&0&0&0&-3&6\sqrt{3}&0&0\\
-18&-\frac{3}{2}&0&-\frac{\sqrt{3}}{2}&0&3&0&-6
\end{array}
\right) . \end{equation}
\ep 
\pf This follows from \re{varphiEqu} with the help of Lemma \ref{KillingLemma} 
\qed 

To compare the above left-invariant metric $g_1$ with the 
metrics obtained from dimensional reduction we need to study 
the automorphism group of the solvable Lie group $L$. Since $L$
is simply connected, we have $\mbox{Aut}(L) \cong \mbox{Aut}(\gl)$.

\subsection{Automorphisms of the solvable algebra \label{Auto}}

In this subsection we determine the automorphism group of the
solvable Lie algebra $\mathfrak{l}$. For the proof we will
use the following dual characterization of automorphisms.
\bp
Given a Lie algebra $\gl$, 
an invertible linear map  $\Lambda: \gl \rightarrow \gl $
is an automorphism if and only if
\begin{equation} \label{DualAut}
d \Lambda^* \theta = \Lambda^* d \theta,
\end{equation}
for all $\theta \in \gl^*$. 
\ep

Recall that given a basis $(T_a)$ of a Lie algebra $\gl$ with
structure constants $c_{ab}^c$, that is $[T_a,T_b]=c_{ab}^c T_c$,
the differential is given in terms of the dual basis $(\theta^a)$
as follows
\[
d\theta^a = - c^a_{bc} \theta^b \wedge \theta^c \;.
\]
In other words, $\Lambda$ is an automorphism if and only if
the dual map $\Omega = \Lambda^*$ satisfies
\begin{equation}
\label{1formautom}
d \Omega(\theta^a) = - c^a_{bc} \Omega (\theta^b) \wedge
\Omega (\theta^c), \;
\end{equation}
for all $a=1 , \ldots, \dim(\gl)$.

For the Iwasawa Lie algebra $\gl$ we can use the basis
$(T_a)=({\cal V}_a)$ defined in (\ref{Vbasis}). The differentials
of the dual basis $(\theta^a)$ are given by (\ref{ExteriorAlg}),
if we put $(\eta^2,\xi_2,\alpha, \beta,\eta^0,\eta^1,\xi_0,\xi_1)= 
(\theta^1, \ldots, \theta^8)$.

We now show the following:

\begin{Th}
%\begin{thm}
The group of automorphisms 
of the solvable Lie algebra $\mathfrak{l}$ 
is given by $\mathrm{Aut}(\mathfrak{l}) = (\mathbbm{Z}_2 \times
\mathbbm{Z}_2) \ltimes \mathrm{Inn}(\mathfrak{l}) $, where 
$\mathrm{Inn}(\mathfrak{l}) \cong L$ denotes the group of inner
automorphisms of $\mathfrak{l}$, and the generators of the cyclic
factors of the group \newline $\mathbbm{Z}_2 \times \mathbbm{Z}_2 \subset 
\mathrm{Aut}(\mathfrak{l})$  act 
by the diagonal matrices 
\[
\mathrm{diag}(-1,1,-1,1,1,-1,-1,1)\quad \mbox{and}\quad
\mathrm{diag}(1,1,1,1,-1, -1, -1, -1), \;
\]
on the Lie algebra $\mathfrak{l}$ with respect to the basis
(\ref{Vbasis}). An explicit parametrization of the group 
$\mathrm{Aut}(\mathfrak{l})$ by 
\[
\{ (a,b,c,d,e,f,g,h) \in \mathbbm{R}^8 | be \not=0 \} \cong
\mathbbm{R}^* \times \mathbbm{R}^* \times \mathbbm{R}^6 ,
\]
is given by the matrix (\ref{MatrixAutl}), which represents the
action of the group element with parameters
$(a,b,c,d,e,f,g,h)$ in the given basis of $\mathfrak{l}$. 
\end{Th}
%\end{thm}

%%%here
\pf
%\begin{proof}
We work with the 1-forms (\ref{DVbasis}), which have exterior 
derivatives (\ref{ExteriorAlg}).

We first note that the six non-zero differentials 
which appear on the right-hand side of (\ref{ExteriorAlg}) are linearly independent.
Hence, the space of closed one-forms $Z^1(\mathfrak{l})$ is spanned by
$\lbrace\xi_2, \beta\rbrace$.

In order to determine all automorphisms $\Lambda$ of $\mathfrak{l}$ we
consider $\Omega=\Lambda^*$ and define coefficients $\Omega^a{}_b$
by $\Omega(\theta^a) = \Omega^a{}_b \theta^b$, such that 
${\cal M} = ({\cal M}_a{}^b)_{a,b} = 
(\Omega^b{}_a)_{a,b}$ is the 
matrix representing $\Omega$ with respect to the basis $(\theta^a)$, and,
hence, is the transpose of the matrix representing $\Lambda$ with respect 
to the basis
$(T_a)$. We then simply work through each of the basis 
1-forms $(\theta^a)$
and determine the coefficients $\Omega^a{}_b$ 
such that \eqref{1formautom} is satisfied. 
It turns out to be easiest to do this in the order 
$\xi_2,\beta,\alpha,\eta^0,\eta^1,\xi_1,\xi_0,\eta^2$.

Since any automorphism preserves $Z^1(\mathfrak{l}) = \mathrm{span}
\{ \xi_2, \beta \}$ we see that
\[
\Omega(\xi_2)=\Omega^2{}_2\xi_2+\Omega^2{}_4\beta \;,\;\;\;
\Omega(\beta)=\Omega^4{}_2\xi_2+\Omega^4{}_4\beta.
\]

We now turn to the 1-form $\alpha$. The automorphism $\Omega$ must satisfy
\[
d\Omega(\alpha)=\frac{1}{\sqrt{3}}\Omega(\alpha)\wedge\Omega(\beta).
\]

Since we have already determined that $\Omega(\beta)$ should be a linear combination of $\lbrace\xi_2, \beta\rbrace$, we deduce that $\Omega(\alpha)$ should only contain terms whose exterior derivative has a $\xi_2$ or $\beta$ in every term. Hence, we require
\[
\Omega(\alpha)=\Omega^3{}_2\xi_2+\Omega^3{}_3\alpha+\Omega^3{}_4\beta+\Omega^3{}_5\eta^0.
\]

We next use the automorphism condition to find algebraic relations between 
the components of $\Omega^3{}_a$ and $\Omega^4{}_a$.
In particular, we have
\[
\Omega^3{}_2\Omega^4{}_4 =\Omega^3{}_4\Omega^4{}_2 \;,
\hspace{5mm} \Omega^3{}_3(\Omega^4{}_4-1) = 0 \;,
\hspace{5mm}\Omega^3{}_3\Omega^4{}_2 =0 \;,
\]
\[\Omega^3{}_5\left(\Omega^4{}_2-\frac{\sqrt{3}}{2}\right) =0 \;,
\hspace{5mm}\Omega^3{}_5\left(\Omega^4{}_4+\frac{3}{2}\right) =0 \;.
\]
From this we see that we can't have \textit{both} $\Omega^3{}_3$ and $\Omega^3{}_5$ being non-zero. 
We consider the case
$\Omega^3{}_3:=b\neq 0, \Omega^3{}_5=0$, which restricts
\[
\Omega^4{}_2=0, \hspace{3mm} \Omega^4{}_4=1, \hspace{3mm} \Omega^3{}_2=0.
\]
It turns out that the other possible choice 
$\Omega^3{}_3=0, \Omega^3{}_5\neq 0$ does not 
give rise to an invertible linear map $\Omega$. The corresponding 
analysis is omitted.

By successively analysing all algebraic relations,
we find the most general automorphism of $\mathfrak{l}$, which depends on 
eight real parameters
\[
\Omega^1{}_2:=a,\,\Omega^3{}_3:=b,\,\Omega^3{}_4:=c,\,\Omega^5{}_2:=d,\,\Omega^5{}_5:=e,\,\Omega^6{}_2:=f,\,\Omega^7{}_2:=g,\,\Omega^8{}_2:=h,
\]
and is given by its action on the basis of 1-forms $(\theta^a)$ as
\begin{align*}\allowdisplaybreaks
\Omega(\xi_2)&=\xi_2, \\
\Omega(\beta)&=\beta, \\
\Omega(\alpha) &= b \alpha+ c\beta, \\
\Omega(\eta^0) &= d\xi_2-\sqrt{3}d\beta +e\eta^0, \\
\Omega(\eta^1) &= f\xi_2+2bd\alpha +\left(2cd-\frac{1}{\sqrt{3}}f\right)\beta -\sqrt{3}ce\eta^0 \\
& \quad +be\eta^1, \\
\Omega(\xi_1)&= h\xi_2-\frac{4}{\sqrt{3}}bf\alpha +\left(\frac{1}{\sqrt{3}}h-\frac{4}{\sqrt{3}}cf\right)\beta -\sqrt{3}c^2e\eta^0 \\
& \quad +2bce\eta^1 +b^2e\xi_1, \\
\Omega(\xi_0)&= g\xi_2 -2bh\alpha +\left(\sqrt{3}g-2ch\right)\beta -c^3e\eta^0 \\
& \quad +\sqrt{3}bc^2e\eta^1 +b^3e\xi_0 +\sqrt{3}b^2ce\xi_1, \\
\Omega(\eta^2)&= b^3e^2\eta^2 +a\xi_2-\left(4bdh+\frac{4}{\sqrt{3}}bf^2\right)\alpha \\
& \quad +\left(2\sqrt{3}dg-4cdh+\frac{2}{\sqrt{3}}fh-\frac{4}{\sqrt{3}}cf^2\right)\beta \\
&\quad  +\left(2\sqrt{3}ceh-2\sqrt{3}c^2ef-2c^3de-2eg\right)\eta^0 \\
&\quad +\left(2\sqrt{3}bc^2de+4bcef-2beh\right)\eta^1 \\
&\quad +2b^3de\xi_0 +\left(2\sqrt{3}b^2cde+2b^2ef\right)\xi_1.
\end{align*}
This eight-parameter family 
describes all automorphisms of the Lie algebra $\mathfrak{l}$. 
We can now read off the matrix ${\cal M}$ 
representing $\Omega = \Lambda^*$ with respect to 
the basis $(\theta^a)$.
%\end{proof}
\begin{equation}
\label{MatrixAutl}
{\cal M}=\left( \begin {array}{cccccccc}
{b}^{3}{e}^{2}&0&0&0&0&0&0&0 \\ 
\noalign{\medskip}a&1&0&0&d&f&g&h\\ 
m_{3,1}
&0&b&0&0&2\,bd&-2\,bh&-4/3\,\sqrt {3} bf\\ 
m_{4,1}
&0&c&1&-\sqrt {3}d&
m_{4,5}&
\sqrt {3}g-2
\,ch&1/3\,\sqrt {3} \left( h-4\,cf \right) \\ 
m_{5,1}
&0&0&0&e&-\sqrt {
3}ce&-{c}^{3}e&-\sqrt {3}{c}^{2}e\\ 
m_{6,1}
&0&0&0&0&be&\sqrt {3}b{c}^{2}e&2
\,bce\\ 
\noalign{\medskip}2\,{b}^{3}de&0&0&0&0&0&{b}^{3}e&0\\
m_{8,1} 
&0&0&0&0
&0&\sqrt {3}{b}^{2}ce&{b}^{2}e\end {array} \right) ,
\end{equation}
where
\[
m_{3,1}= -4\,b \left( 
dh+1/3\,\sqrt {3}{f}^{2} \right), \;\;\;
m_{4,1} = 2\,\sqrt {3}dg-4\,cdh+2/3\,\sqrt {3}fh-4/3\, \sqrt {3}c{f}^{2},
\]
\[
m_{4,5} =  2\,cd-1/3\,\sqrt {3}f \;,\;\;\;
m_{5,1} = 2\, \sqrt {3}ceh-2\,\sqrt {3}{c}^{2}ef-2\,{c}^{3}de-2\,eg,
\]
\[
m_{6,1} = 2\,b \left( 
\sqrt {3}{c}^{2}de+2\,cef-eh \right) \;,\;\;\;
m_{8,1} = 2\,{b}^{2} \left( \sqrt {3}cde+ef \right) .
\]
Note that the matrix ${\cal M}$ satisfies the equation
$\Lambda(T_a) = {\cal M}_a{}^b T_b$. 

Since $\det(M) = b^{10} e^6$ is not allowed to be zero, we conclude that
$b\not=0$ and $e\not=0$, which decomposes the eight-parameter family 
into four connected components. Notice that the matrices 
${\cal M}$ such that $a=c=d=f=g=h=0$ and $b, e \in \{ \pm 1\}$
form a subgroup of $\mathrm{Aut}(\mathfrak{l})$ isomorphic to 
$\mathbbm{Z}_2 \times  \mathbbm{Z}_2$. Its action on $\mathfrak{l}$
is diagonal as indicated in the theorem, 
and can be read off from (\ref{MatrixAutl}).  
\qed

\subsection{Identifying the open orbit corresponding to Time-\\Space reduction}\label{IdTS}

Under automorphisms, the Gram matrix ${\cal G}_1$ given 
by  (\ref{Gramlprime}) transforms according to
\[
{\cal G}_1 \mapsto {\cal M}{\cal G}_1{\cal M}^T \;,
\]
where ${\cal M}$ is the matrix 
(\ref{MatrixAutl}) representing the dual of a general automorphism
of the Iwasawa algebra $\mathfrak{l}'$.
We now  impose that the transformed Gram matrix is diagonal up to scale.
The related calculations can be easily performed 
using Maple. By imposing successively the 
vanishing of off-diagonal entries of the transformed Gram matrix, one obtains
constraints on the eight parameters of the automorphism. 
The parameters
have to take the values
\[
d=-\frac{1}{\sqrt{3}} ,\;\;\;
f=-\frac{1}{2},\;\;\;
h=g=0 ,\;\;\;
a=-\frac{1}{6},\;\;\;
c=-\frac{1}{2},\;\;\;
b=\pm\frac{1}{2},\;\;\;
e=\pm \frac{2}{\sqrt{3}} .
\]
This shows that there is a unique inner automorphism ($b,e>0$) which
diagonalizes the Gram matrix, and as well a unique such automorphism
in each component of $\mbox{Aut}(\gl)$.
The diagonalized Gram matrix is in all cases
\begin{equation}
\label{Diaglp}
{\cal G}_1^{\rm diag} = \frac{1}{2} \mbox{diag}
(-1,1,-1,1,-1,1,1-1) \;.
\end{equation}
This agrees, up to an overall (positive) factor, with the metric
(\ref{Metric1}), (\ref{signature})
of the scalar manifold $M^{(TS)}$ obtained by TS reduction
($\epsilon_1=1, \epsilon_2=-1$). We have therefore shown:
\bp 
The left-invariant metric $g^{(TS)}=g^{(1,-1)}$ on $L$ obtained by dimensional
reduction of five-dimensional minimal 
pure supergravity is related by a unique inner automorphism combined with 
a re-scaling by a factor of $\frac{1}{2}$
to the left-invariant metric
$g_1$ on $L$ obtained from the open orbit $M_1 = L' \cdot o \subset S=
G/G_{ev}$, where 
$L'=C_a(L)$, $a=\exp(e^1 + e^1_3)\in G$, 
is the Iwasawa subgroup constructed in Subsection \ref{IwaSub}.
\ep 

\subsection{Identifying the open orbit corresponding to Space-Time reduction} \label{IdST}

Next, we look for another $a\in G$ such that the
Iwasawa subalgebra $\mathfrak{l}'' = \mbox{Ad}_a(\gl)
\subset \gg$,
is transversal to $\gg_{ev}$, and hence gives rise to a second
open orbit $M_2=L'' \cdot o \subset S$, where $L'' = \exp(\gl'')$.
The aim is to match $M_2$ with $M^{(ST)}$, up to a covering, using
again an inner automorphism of $L$ to relate the corresponding 
left-invariant metrics
$g_2 = \varphi_2^* g_S$ and $g^{(ST)}$ on $L$. 
Here $\varphi_2: L \rightarrow M_2$ is the covering
$x\mapsto C_a(x) \cdot o$. 
This procedure involves choosing 
$\xi \in \mathfrak{g}$ such that $a=\exp(\xi)$ has the desired properties.
Investigating candidates for $\xi$ is tedious but manageable using Maple.
Otherwise we follow the same steps as for $\gl'$.

We use the following basis of $\mathfrak{g}$:
\[
b_1 = e_1^1 -e_2^2 ,\;\;\;
b_2 = e_2^2 - e_3^3 ,\;\;\;
b_3 = e_1^2 ,\;\;\;
b_4 = e_1^3 ,\;\;\;
b_5 = e_2^3 ,\;\;\;
b_6 = e_2^1  ,\;\;\;
b_7 = e_3^1 ,\;\;\;
\]
\[
b_8 = e_3^2 \;,\;\;\;
b_9 = e_1 \;,\;\;\;
b_{10} = e_2 \;,\;\;\;
b_{11} = e_3 \;,\;\;\;
b_{12} = e^1 \;,\;\;\;
b_{13} = e^2 \;,\;\;\;
b_{14} = e^3 .
\]
Note that $\mathfrak{l}=\mbox{span} \{ b_1, b_2, b_3, b_4, b_5, b_9, 
b_{13}, b_{14} \}$, with the relation to the basis $({\cal V}_b)$
given by (\ref{V-basis}).  
We take $\xi=e_3^2 + e^1$ and compute $X' = \mbox{Ad}_a X$,
where $a = \exp \xi$, 
for all basis elements $X=b_m$ of $\mathfrak{l}$:
\[
b'_1 = b_1 - b_8 + b_{12} \;,\;\;\;
b'_2 = b_2 + 2 b_8 \;,\;\;\;
b'_3 = b_3-b_7+b_{11} +b_{13} \;,\;\;\;
\]
\[
b'_4 = - b_3 +b_4 + b_6 + b_7 -b_{10} - b_{11} -b_{13} + b_{14} 
\;,\;\;\;
\]
\[
b'_{5} = - b_2 + b_5 - b_8 \;,\;\;\;
b'_9 = -2 b_1 - b_2 + b_9 - b_{12} \;,\;\;\;
\]
\[
b'_{13} = -3 b_7 + 2 b_{11} + b_{13} \;,\;\;\;
b'_{14} = 3 b_6 + 3 b_7 -2 b_{10} -2 b_{11} - b_{13} + b_{14} \;.
\]

As before we denote by $\varphi$ the composition 
$\pi \circ \mbox{Ad}_a$ where $\pi: \gg \rightarrow \gg_{odd}$
is the projection along $\gg_{ev}$. Using the above formulae
we apply $\varphi$ to the basis elements 
${\cal V}_b$ and express the result in the basis
$(f_1, \ldots, f_8):=
( b_9, b_{10}, b_{12}, b_{13}, b_4, b_5, b_7, b_8)$ 
of $\mathfrak{g}_{\rm odd}$. 
For example
\[
\varphi ({\cal V}_1) = -3 \varphi (b_4) = -3 b_4 -3 b_7 + 3 b_{10}
+ 3 b_{13} \;.
\]
The result is summarized by the matrix $A$, 
which is the transpose of the matrix representing $\varphi: \gl
\rightarrow \gg_{odd}$ with respect to the bases $({\cal V}_b)$ and 
$(f_b)$, that is
$\varphi({\cal V}_b) = A_{bc} f_c$:
\[
A = 
 \left( \begin {array}{cccccccc} 0&3&0&3&-3&0&-3&0
\\ \noalign{\medskip}0&0&\frac{1}{2}&0&0&0&0&\frac{1}{2}\\ \noalign{\medskip}0&0&0&\frac{1}{\sqrt{3}}&0&0&-\sqrt {3}&0\\ \noalign{\medskip}0&0&\frac{1}{2\sqrt{3}}&0&0
&0&0&-\frac{\sqrt{3}}{2}\\ \noalign{\medskip}0&0&0&0&0&\sqrt {3}&0&-\sqrt 
{3}\\ \noalign{\medskip}0&-2&0&-1&0&0&3&0\\ \noalign{\medskip}0&0&0&-
\sqrt {3}&0&0&\sqrt {3}&0\\ \noalign{\medskip}-1&0&1&0&0&0&0&0
\end {array} \right).
\]
One checks that $\det(A) = -12 \not=0$, and therefore the vectors 
$\varphi ({\cal V}_b)$ are linearly independent, and $\mathfrak{l}'' =
\mbox{span} \{ {\cal V}_b \} \simeq \mathfrak{g}_{\rm odd}$ is
transversal. The Gram matrix ${\cal G}_2$ of the scalar product
$\langle \cdot , \cdot \rangle_2 = \varphi^* \langle \cdot , \cdot \rangle_B$ 
on $\gl$ with respect to the basis $({\cal V}_b)$ is given by 
\[
{\cal G}_2 = A {\cal G} A^T ,
\]
where ${\cal G}$ is the Gram matrix of the scalar product $\langle \cdot,
\cdot \rangle_B$ on $\gg_{odd}$ with respect to the basis $(f_b)$, 
as computed in Lemma \ref{KillingLemma}. The resulting matrix is 
\begin{equation}
\label{Gramlpp}
{\cal G}_2= \left( \begin {array}{cccccccc} 72&0&6\,\sqrt {3}&0&0&-36&-12\,\sqrt 
{3}&0\\ \noalign{\medskip}0&0&0&0&\frac{\sqrt{3}}{2}&0&0&-\frac{3}{2}
\\ \noalign{\medskip}6\,\sqrt {3}&0&0&0&0&-2\,\sqrt {3}&0&0
\\ \noalign{\medskip}0&0&0&0&-\frac{3}{2}&0&0&-\frac{\sqrt{3}}{2}
\\ \noalign{\medskip}0&\frac{\sqrt{3}}{2}&0&-\frac{3}{2}&-6&0&0&0
\\ \noalign{\medskip}-36&0&-2\,\sqrt {3}&0&0&12&6\,\sqrt {3}&0
\\ \noalign{\medskip}-12\,\sqrt {3}&0&0&0&0&6\,\sqrt {3}&0&0
\\ \noalign{\medskip}0&-\frac{3}{2}&0&-\frac{\sqrt{3}}{2}&0&0&0&-6\end {array}
 \right) \;. 
\end{equation}
%%%
Now we apply a general automorphism of $\mathfrak{l}$ with matrix ${\cal M}$
as in (\ref{MatrixAutl})
and impose that ${\cal M} {\cal G}_2 {\cal M}^T$ is diagonal up to scale. This
leads to the following constraints on the parameters of ${\cal M}$:
\[
f = 0 ,\;\;\;
%\frac{c}{4} -\sqrt{3} cd ,\;\;\;
d= \frac{1}{12} \sqrt{3} ,\;\;\;
h=-\frac{1}{4} ,\;\;\;
g= 0 ,\;\;\;
a=0 ,\;\;\;
c=0 ,\;\;\;
b=\pm 1 ,\;\;\; 
e = \pm \frac{1}{2\sqrt{3}}  .
\]
Thus there is again a unique inner  automorphism diagonalizing the
Gram matrix, and precisely one such automorphism in each component
of $\mbox{Aut}(\gl)$. The diagonalized 
Gram matrix is in all cases 
\begin{equation}
\label{Diaglpp}
{\cal G}_2^{\rm diag} = \frac{1}{2} 
\mbox{diag}(1,1,1,1,-1,-1,-1,-1) \;,
\end{equation}
which agrees, up to an overall (positive) scale factor with 
the metric $g^{(ST)}$ 
(\ref{Metric1}), (\ref{signature})
of the scalar manifold $M^{(ST)}$ obtained by ST reduction,
($\epsilon_1=-1, \epsilon_2=1$). Thus we have shown: 
\bp
The left-invariant metric $g^{(ST)}=g^{(-1,1)}$ on $L$ obtained by dimensional
reduction of five-dimensional minimal 
pure supergravity is related by a unique inner automorphism combined with
a re-scaling by a factor of $\frac{1}{2}$
to the left-invariant metric
$g_2$ on $L$ obtained from the open orbit $M_2 = L'' \cdot o \subset S=
G/G_{ev}$, where 
$L''=C_a(L)$, $a=\exp(e^2_3 + e^1)\in G$, 
is the Iwasawa subgroup constructed above.
\ep

%%%
We also have:
\bp
\label{PropNonAut}
The left-invariant metrics $g^{(TS)}$ and $g^{(ST)}$ (equivalently $g_1$ and $g_2$)
on $L$ are not related by any automorphism of $L$. 
\ep
\pf
This can be proven by showing that an automorphism 
cannot transform 
the diagonal Gram matrix ${\cal G}_1^{\rm diag}$ (\ref{Diaglp}) 
of $g^{(TS)}$  to the diagonal Gram matrix ${\cal G}_2^{\rm diag}$ 
(\ref{Diaglpp}) of $g^{(ST)}$.  If such an automorphism
existed, then there would exist a matrix ${\cal M}$ of the form 
(\ref{MatrixAutl}) such that 
${\cal M} {\cal G}_1^{\rm diag} {\cal M}^T = 
{\cal G}_2^{\rm diag}$.
 Analysing the conditions imposed on the parameters, one finds
$a=d=f=g=h=0$, 
which in turn implies (amongst other things) that 
$\Omega(\eta^2)=b^3e^2\eta^2$. 
%Hence, the ``1-1'' component of 
%${\cal  {\cal G}''_{\rm diag}  \Omega=
%{\cal G}'_{\rm diag}$ requires 
%that $C_3^6 E_5^4=-1$ which is impossible.
Hence we see that we need $b^6 e^4=-1$ which is impossible.

Alternatively, this follows from the uniqueness of the 
diagonalization of the Gram matrices ${\cal G}_1$ and
${\cal G}_2$ by automorphisms, which was observed above.
\qed

In the next section we will investigate the geometry of the manifolds
$(L,g_1)$ and $(L,g_2)$ 
more closely.

%%%%%%%%%%%%%%%%%%%%%%%%%%%%%%%%%%%%%%%%%%%%%%%%%%%%%%%%%%%%%%%%%%%%

\section{\sloppy Geometric structures on the Iwasawa subgroup of $G_{2(2)}$} \label{GeomStruct}

%%%Overview?%%%%%%%%%%%%%%%%%%%%%%%%%%%%%%%%%%%%%%%%%%%%%%%%%%%%%%%%

We now explore the geometrical structures carried by the
Lie algebra $\mathfrak{l}$ of the Iwasawa subgroup $L$,
equipped with the three metrics related to dimensional
reductions of minimal pure five-dimensional supergravity,
and find explicit expressions for the (para-)quaternionic
structures, the Levi-Civita connection and the curvature.

Recall that the Lie algebra $\gl$ is equipped with the
basis $({\cal V}_a)$ defined in (\ref{V-basis}) and the
left-invariant metrics $g_1, g_2$ with the Gram matrices
${\cal G}_1$, ${\cal G}_2$ given in (\ref{Gramlprime})
and (\ref{Gramlpp}). As observed above there exists in both
cases a 
unique inner automorphism of $\gl$ transforming the above
Gram matrices into diagonal forms given by (\ref{Diaglp}),
(\ref{Diaglpp}). Let us denote in both cases 
by $(T_a) = (V_2, U^2, A, B, V_0, V_1, U^0, U^1 )$ 
the, up to scale,
orthonormal basis which
corresponds to the basis $({\cal V}_a)$ under this unique
inner automorphism.  
For convenience, in this section we work with rescaled
metrics on $L$ and corresponding scalar products on $\gl$, 
for which $(T_a)$ is an orthonormal basis. For completeness
we also consider the case of SS reduction which corresponds
to a positive definite metric on $L$. Thus the 
Gram matrices with respect to the basis $(T_a)$ for the
three cases are:
\begin{equation}
\label{GramON}
{\cal G}^{\rm diag} = (-\epsilon_1, 1, -\epsilon_1,1, -\epsilon, -\epsilon_2,
-\epsilon_2, -\epsilon) \;.
\end{equation}
The corresponding metric on $L$ is denoted $\bar{g}=
\bar{g}^{(\epsilon_1,\epsilon_2)}$ and the associated 
scalar product on $\mathfrak{l}$ by $\langle \cdot, \cdot \rangle =
\langle \cdot, \cdot \rangle^{\epsilon_1, \epsilon_2}$.

Summarizing we are given the Lie algebra $\mathfrak{l}$ with the
basis 
\begin{equation}
\label{T-basis}
(T_a) = (V_2, U^2, A, B, V_0, V_1, U^0, U^1 ),
\end{equation}
with structure constants (\ref{Iwasawa1}) and a pseudo-Euclidean
scalar product $\langle \cdot, \cdot \rangle$ defined by the
Gram matrix (\ref{GramON}) with respect to the basis $(T_a)$. 
We now state the main results which will be proved in this
section.

We first define the following skew-symmetric endomorphisms.
\begin{eqnarray}
J_1 &=& \epsilon_2 U^2 \wedge V_2 - B \wedge A 
+ \epsilon \frac{\sqrt{3}}{2} U^1 \wedge U^0 
- \epsilon_2 \frac{1}{2} U^1 \wedge V_1 \nonumber\\
&& + \epsilon_2 \frac{1}{2} U^0 \wedge V_0
+ \epsilon \frac{\sqrt{3}}{2} V_1 \wedge V_0  \;, \label{J1}\\
J_2 &=& \epsilon_2 \frac{\sqrt{3}}{2} U^1 \wedge V_2 
+ \epsilon \frac{1}{2} V_0 \wedge V_2
- \frac{1}{2} U^0 \wedge U^2 
- \epsilon_1 \frac{\sqrt{3}}{2} V_1 \wedge U^2 \nonumber \\
&& - \frac{1}{2} U^1 \wedge A
- \epsilon_1 \frac{\sqrt{3}}{2} V_0 \wedge A 
- \frac{\sqrt{3}}{2} U^0 \wedge B 
+ \epsilon_1 \frac{1}{2} V_1 \wedge B \;, \\
J_3 &=& \epsilon_2 \frac{1}{2} U^0 \wedge V_2
- \epsilon\frac{\sqrt{3}}{2} V_1 \wedge V_2 
- \epsilon_1 \frac{\sqrt{3}}{2} U^1 \wedge U^2 
+ \frac{1}{2} V_0 \wedge U^2  \nonumber \\
&& 
- \frac{\sqrt{3}}{2} U^0 \wedge A 
+ \epsilon_1 \frac{1}{2} V_1 \wedge A 
- \epsilon_1 \frac{1}{2} U^1 \wedge B 
- \frac{\sqrt{3}}{2} V_0 \wedge B \;, \label{J3}\\
\tilde{J}_1 &=& - \epsilon_2 U^2 \wedge V_2 - B \wedge A 
+ \epsilon \frac{\sqrt{3}}{2} U^1 \wedge U^0 
- \epsilon_2 \frac{1}{2} U^1 \wedge V_1 \nonumber \\
&& 
+ \epsilon_2 \frac{1}{2} U^0 \wedge V_0
+ \epsilon \frac{\sqrt{3}}{2} V_1 \wedge V_0 \;. 
\label{J1tilde}
\end{eqnarray}

Here we use the following standard identification of bi-vectors with
skew-symmetric endomorphisms: 
\begin{equation}
\label{skew_end}
(u \wedge v)(w) = u \langle v , w\rangle - \langle u, w \rangle v \;,\;\;\;
u,v,w, \in \mathfrak{l} \;.
\end{equation}

\bp 
The endomorphisms $J_\alpha$ of $\mathfrak{l}$
are pairwise anti-commuting and satisfy the following
relations\footnote{Recall that $\epsilon = - \epsilon_1 \epsilon_2$. }:
\[
(J_1)^2 = \epsilon_1 \mbox{Id} \;,\;\;\;
(J_2)^2 = \epsilon_2 \mbox{Id} \;,\;\;\;
(J_3)^2 = \epsilon_3  \mbox{Id} := \epsilon \mbox{Id} \;,\;\;\;
J_3 = J_1 J_2 \;.
\]
\ep

\pf
This follows by direct calculation.
\qed

Notice that the endomorphisms $J_\alpha$ define 
left-invariant skew-symmetric almost $\epsilon_\alpha$-complex\footnote{By the terminology ``$\epsilon$-complex'', ``$\epsilon$-quaternionic'', etc.\ we mean ``complex'', ``quaternionic'', etc.\ if $\epsilon=-1$ and ``para-complex'', ``para-quaternionic'', etc.\ if $\epsilon=1$.} 
structures on the Lie group $L$, which will be denoted
by the same symbols. We put $Q:=\mbox{span}\{J_\alpha | \alpha=1,2,3 \}$.

\bt
\label{Thm2}
$(L,\bar{g},Q)$ is an $\epsilon$-quaternionic-K\"ahler manifold 
with left-invariant $\epsilon$-quaternionic structure $Q$, 
and
$J_1, \tilde{J}_1$ are integrable left-invariant skew-symme\-tric 
$\epsilon_1$-complex structures on $(L,\bar{g})$. 
\et

\pf
The integrability of the structures $J_1, \tilde{J}_1$ is proven 
by computation of the Nijenhuis tensor\footnote{
Notice that by the Newlander-Nirenberg theorem and the 
Frobenius theorem, respectively, the vanishing of the Nijenhuis
tensor $N_J$ of an almost $\epsilon$-complex structure $J$ on a smooth 
manifold $M$ implies that $J$ defines on $M$ the structure of a
complex, respectively para-complex, manifold.} 
\begin{equation}
\label{Nijenhuis}
N_J (X,Y) = - J^2 [X,Y] + J[JX,Y] + J[X,JY] - [JX,JY]  = 0 \;,
\end{equation}
for $J=J_1$ and $J=\tilde{J}_1$ using the formulae
(\ref{Iwasawa1}), (\ref{J1}) and (\ref{J1tilde}). One also finds that $N_{J_2}, N_{J_3}$ do
not vanish.
To prove the $\epsilon$-quaternionic-K\"ahler property we need to 
check that $Q$ is parallel.  The explicit expression for the Levi-Civita 
connection is given in formula (\ref{Connection}) in Subsection
\ref{LC}. Using this formula, it is checked in Proposition \ref{PropNablaLC}.
that $Q$ is parallel.
\qed

The curvature tensor of the $\epsilon$-quaternionic-K\"ahler manifold
$(L,\bar{g},Q)$ 
is given in formula (\ref{Curvature}) in Subsection \ref{SectCurvature}.
Based on the formulae for the Levi-Civita connection and its 
curvature we have verified by explicit calculation that the
curvature tensor is parallel. This provides a second, independent
proof of the fact, established in Section \ref{SectRealization},
that the manifold $(L,\bar{g})$ is locally symmetric.

For $\epsilon_1=\epsilon_2=-1$ Theorem \ref{Thm2} recovers
Alekseevsky's description \cite{MR0402649}
of the symmetric quaternionic-K\"ahler manifold
of non-compact type $G_{2(2)}/SO_4$ as a solvable Lie group $L$
endowed with a left-invariant quaternionic-K\"ahler structure.
For completeness we include in Subsection \ref{Sect_Q-reps}
a discussion relating our approach
with Alekseevsky's description in terms of representations of
K\"ahlerian Lie algebras.

\subsection{Computation of the Levi Civita connection \label{LC}}

To compute the Levi-Civita connection of a pseudo-Riemannian
metric $g$, we use the Koszul
formula
\begin{eqnarray}
2 g(\nabla_X Y, Z) &=& X  g(Y,Z) + Y  g(X,Z) 
- Z  g(X,Y) \nonumber \\
&& + g([X,Y],Z) - g(X,[Y,Z]) 
- g(Y,[X,Z]), \nonumber
\end{eqnarray}
where $X,Y,Z$ are vector fields\footnote{When taking
$X,Y,Z$ to be coordinate vector fields (and hence to commute), the
last three terms on the right hand side 
vanish and one recovers the usual formula
for the Levi-Civita connection in terms of Christoffel symbols.}. 
For a left-invariant metric on a Lie group $L$ the vector fields $X,Y,Z$
can be taken to be left-invariant and therefore correspond
to vectors in the Lie algebra $\mathfrak{l}$, in which case the first three 
terms on the right hand side vanish. 
The computation of the
Levi-Civita connection is thus reduced to computing commutators and
scalar products of vectors in  $\mathfrak{l}$. 
Notice that the covariant derivative $\nabla_X$ acts 
on $\mathfrak{l}$ as an endomorphism, which satisfies
\[
T(X,Y) = \nabla_X Y - \nabla_Y X - [X,Y] = 0 \;,\;\;\;\forall X,Y \in 
\mathfrak{l} \;,
\]
and which is skew as 
a consequence of the metric compatibility of the Levi-Civita connection.
Therefore we can express $\nabla_X$ as a wedge product of generators,
using the convention (\ref{skew_end}).

Using the commutators (\ref{Iwasawa1}) in the solvable Lie algebra
$\mathfrak{l}$ 
together with the fact that the generators $T_a$ 
form an orthonormal
basis (\ref{T-basis}) with the Gram matrix (\ref{GramON})
it is straightforward to obtain the following result.

\bp
\label{PropLC}
The Levi-Civita connection $\nabla$ of $(L,\bar{g})$ is given by:
\begin{eqnarray}\label{Connection}
\nabla_{V_{2}} &=& U^{2}\wedge V_{2}+\frac{1}{2}U^0\wedge V_0 +\frac{1}{2}U^1\wedge V_1,  \\
\nabla_{U^{2}} &=& 0, \nonumber \\
\nabla_A &=& \frac{1}{\sqrt{3}}B\wedge A+\frac{1}{2}\epsilon V_0\wedge V_1 +\frac{1}{2}\epsilon U^0\wedge U^1+ \frac{1}{\sqrt{3}}\epsilon_2 U^1\wedge V_1, \nonumber \\
\nabla_B &=& 0, \nonumber \\
\nabla_{V_0}&=& -\frac{1}{2}V_0\wedge U^{2}-\frac{1}{2}\epsilon_2 U^0\wedge V_{2}
+\frac{\sqrt{3}}{2}V_0\wedge B+\frac{1}{2}\epsilon_1 V_1\wedge A, \nonumber \\
\nabla_{V_1} &=& -\frac{1}{2}V_1\wedge U^{2}-\frac{1}{2}\epsilon U^1\wedge V_{2}
+\frac{1}{2\sqrt{3}}V_1\wedge B \nonumber \\
& &-\frac{1}{2}V_0\wedge A -\frac{1}{\sqrt{3}}\epsilon_1 U^1\wedge A, \nonumber \\
\nabla_{U^0} &=& -\frac{1}{2}U^0\wedge U^{2}+\frac{1}{2}\epsilon V_0\wedge V_{2}
-\frac{\sqrt{3}}{2}U^0\wedge B +\frac{1}{2}U^1\wedge A, \nonumber \\
\nabla_{U^1} &=&-\frac{1}{2}U^1\wedge U^{2}+\frac{1}{2}\epsilon_2 V_1\wedge V_{2} 
-\frac{1}{2\sqrt{3}}U^1\wedge B \nonumber \\
& &-\frac{1}{2}\epsilon_1 U^0\wedge A+\frac{1}{\sqrt{3}}V_1\wedge A. \nonumber
\end{eqnarray}
\ep
For illustration, the first line of the above formula 
is equivalent to:
\[
\nabla_{V_2} V_2 = - \epsilon_1 U^2 \;,\;\;\;
\nabla_{V_2} U_2 = - V_2 \;,\;\;\;
\nabla_{V_2} A = 0 \;,\;\;\;
\nabla_{V_2} B = 0 \;.
\]
\[
\nabla_{V_2} V_0 = - \frac{1}{2} \epsilon U^0 \;,\;\;\;
\nabla_{V_2} U^0 = \frac{1}{2}\epsilon_2 V_0 \;,\;\;\;
\nabla_{V_2} V_1 = - \frac{1}{2} \epsilon_2 U^1 \;,\;\;\;
\nabla_{V_2} U^1 = \frac{1}{2} \epsilon V_1 \;.\;\;\;
\]

%%%%%%%%%%%%%%%%%%%%%%%%%%%%%

The covariant derivatives of the 
structures $J_{\alpha}$ can now be computed by taking
commutators between the corresponding skew endomorphisms
$\nabla_X$ and $J_\alpha$ for all 
$X \in \mathfrak{l}$. We note the following useful formula:
\[
[X\wedge Y, Z \wedge W ] = (X\wedge W) \langle Y,Z \rangle
+ (Y\wedge Z) \langle X,W \rangle
- (X\wedge Z) \langle Y, W \rangle
- (Y \wedge W) \langle X, Z \rangle \;.
\]
Using Proposition \ref{PropLC} and the explicit expressions
for the structures $J_\alpha$ given in (\ref{J1})--(\ref{J3}),
we obtain:
\bp
\label{PropNablaLC}
\begin{eqnarray}\label{epsilonQKstructure}
\left[\nabla_X,J_1\right]&=&\hat{\alpha}(X)J_2+\hat{\beta}(X)J_3, \nonumber \\
\left[\nabla_X,J_2\right]&=&\epsilon \hat{\alpha}(X)J_1
+\hat{\gamma}(X)J_3, \\
\left[\nabla_X,J_3\right]&=&\epsilon_2\hat{\beta}(X)J_1+\epsilon_1\hat{\gamma}(X)J_2, \nonumber
\end{eqnarray}
where $\hat{\alpha},\hat{\beta},\hat{\gamma}\in\mathfrak{l}^*$ 
are one-forms,
which are related to the dual basis
\eqref{1forms} of the basis $(T_a)$ of $\mathfrak{l}$  by
\begin{equation}
\hat{\alpha}=-\frac{1}{2}\eta^0-\frac{\sqrt{3}}{2}\xi_1, \hspace{3mm} 
\hat{\beta}=-\epsilon_1\frac{\sqrt{3}}{2}\eta^1-\frac{1}{2}\xi_0, \hspace{3mm} \hat{\gamma}=\epsilon_2\frac{1}{2}\eta^2-\frac{\sqrt{3}}{2}\alpha.
\end{equation}
\ep
This shows that the $\epsilon$-quaternionic structure $Q$ is parallel.

\subsection{Curvature}\label{SectCurvature}

Given the Levi-Civita connection it is straightforward
to compute the curvature by the defining formula
\[
R(X,Y) = [ \nabla_X, \nabla_Y] - \nabla_{[X,Y]} \;,
\]
where $X,Y$ are vector fields. In our setting we take $X,Y$ to 
be left-invariant vector fields on the Lie group $L$ and identify
them with elements of $\mathfrak{l}$. Then 
$R(X,Y)$ is considered as a skew-symmetric endomorphism of $\mathfrak{l}$.

\bp
The curvature endomorphisms of $(L,\bar{g})$ are given by:
\begin{eqnarray}\label{Curvature}
& & R(U^{2},V_{2}) = -\nabla_{V_{2}}, \quad R(U^{2},B)=R(U^{2},A)=0, \nonumber \\
& & R(U^{2},V_0)= -\frac{1}{2}\nabla_{V_0}, \quad R(U^{2},V_1)=-\frac{1}{2}\nabla_{V_1}, \quad R(U^{2},U^0)=-\frac{1}{2}\nabla_{U^0}, \nonumber \\  & & R(U^{2},U^1)= -\frac{1}{2}\nabla_{U^1}, \nonumber \\
& & R(V_{2},B)= R(V_{2},A)=0, \quad R(V_{2},V_0)=\frac{1}{2}\epsilon\nabla_{U^0}, \nonumber \\
& & R(V_{2},V_1)= \frac{1}{2}\epsilon_2\nabla_{U^1}, \quad R(V_{2},U^0) =-\frac{1}{2}\epsilon_2\nabla_{V_0},
\quad R(V_{2},U^1)= -\frac{1}{2}\epsilon\nabla_{V_1}, \nonumber \\
& & R(B,A)= -\frac{1}{\sqrt{3}}\nabla_A , \quad R(B,V_0)=\frac{\sqrt{3}}{2}\nabla_{V_0}, \quad R(B,V_1)=\frac{1}{2\sqrt{3}}\nabla_{V_1}, \nonumber \\
& & R(B,U^0)=-\frac{\sqrt{3}}{2}\nabla_{U^0}, \quad R(B,U^1)=-\frac{1}{2\sqrt{3}}\nabla_{U^1}, \nonumber \\
& & R(A,V_0)= -\frac{1}{2}\nabla_{V_1}, \quad R(A,V_1)=\frac{1}{2}\epsilon_1\nabla_{V_0}+\frac{1}{\sqrt{3}}\nabla_{U^1}, \nonumber \\
& & R(A,U^0)= -\frac{1}{2}\epsilon_1\nabla_{U^1}, \quad R(A,U^1)=\frac{1}{2}\nabla_{U^0}-\frac{1}{\sqrt{3}}\epsilon_1\nabla_{V_1}, \nonumber \\
& & R(V_0,V_1)= -\frac{1}{2}\epsilon\nabla_{A}, \quad R(V_0,U^0)=\frac{1}{2}\nabla_{V_{2}}, \quad R(V_0,U^1)=0, \nonumber \\
& & R(V_1,U^0)= 0, \quad R(V_1,U^1)=\frac{1}{2}\nabla_{V_{2}}+\frac{1}{\sqrt{3}}\epsilon_2\nabla_{A}, \quad R(U^0,U^1)= -\frac{1}{2}\epsilon\nabla_{A}. \nonumber \\
\end{eqnarray}
\ep

From a tedious but straightforward calculation we deduce:
\bc
The pseudo-Riemannian manifold $(L,\bar{g})$ is locally symmetric:
\[ \nabla R = 0. \]
\ec

\subsection{Relation with $Q$-representations of K\"{a}hlerian Lie algebras in the Riemannian case \label{Sect_Q-reps}}

%{\bf Paul, please check, and add missing structures! PD: Done}

In this section we explain the relation between our construction
and Alekseevsky's classification
of left-invariant quaternionic-K\"ahler
structures on solvable Lie groups 
using the so-called $Q$-representations of K\"ahlerian 
Lie algebras \cite{MR0402649}. In the Riemannian case of 
our construction, $\epsilon_1=\epsilon_2=-1$, we can make
the following unitary basis transformation of the Lie 
algebra:
\[
G_0 := V_2 \;,\;\;\;
H_0 := U^2 \;,\;\;\;
G_1 := A \;,\;\;\;
H_1 := B \;,\;\;\;
\]
\[
\tilde{G}_0 := - \frac{\sqrt{3}}{2} U^1 - \frac{1}{2} V_0 \;,\;\;\;
\tilde{G}_1 := - \frac{1}{2} U^1 + \frac{\sqrt{3}}{2} V_0 \;,\;\;\;
\]
\[
\tilde{H}_0 := - \frac{1}{2} U^0 + \frac{\sqrt{3}}{2} V_1 \;,\;\;\;
\tilde{H}_1 := - \frac{\sqrt{3}}{2} U^0 - \frac{1}{2} V_1 \;.
\]
In this new orthonormal basis, the complex structures take the
following simple form:
\begin{eqnarray}
J_1 &=& - H_0 \wedge G_0 - H_1 \wedge G_1 
+ \tilde{H}_0 \wedge \tilde{G}_0  + \tilde{H}_1 \wedge \tilde{G}_1, 
\nonumber \\
J_2 &=& \tilde{G}_0 \wedge G_0 
+ \tilde{H}_0 \wedge H_0
+ \tilde{G}_1 \wedge G_1  
+ \tilde{H}_1 \wedge H_1, \nonumber \\
J_3 &=& -\tilde{G}_0\wedge H_0
+ \tilde{H}_0 \wedge G_0
- \tilde{G}_1 \wedge H_1  
+ \tilde{H}_1 \wedge G_1, \nonumber \\
\tilde{J}_1 &=& H_0 \wedge G_0 
- H_1 \wedge G_1 
+ \tilde{H}_0 \wedge \tilde{G}_0  
+ \tilde{H}_1 \wedge \tilde{G}_1.  %\\ \nonumber 
\end{eqnarray}
In Alekseevsky's construction, when applied to the special case
of $(L,\bar{g})\cong G_{2(2)}/SO_4$,
one starts with the sum of two elementary K\"ahlerian Lie 
algebras
\[
\mathfrak{u} = \mathfrak{f}_0 \oplus \mathfrak{f}_1,
\]
where
\[
\gf_0  
=\mbox{span}\{G_0, H_0 \} \;,\;\;\;
%= \mbox{span} \{V_2, U^2\} \;,\;\;\;
[H_0, G_0 ] = G_0 \;,
%[U^2,V_2]=V_2 \;,
\]
\[
\gf_1 
= \mbox{span}\{G_1, H_1 \} \;,\;\;\;
%= \mbox{span}\{A,B\}\;,\;\;\;
[H_1, G_1] = \frac{1}{\sqrt{3}} G_1 \;.
%[B,A] = \frac{1}{\sqrt{3}} A \;.
\]
The corresponding Lie group $U$ acts simply transitively on the
product of two complex hyperbolic lines with curvatures
$-1$ and $-\frac{1}{3}$, respectively. 
The latter 
is the projective special K\"ahler manifold obtained by 
applying the local $r$-map to a zero-dimensional manifold.
The symmetric space corresponding to the complex
hyperbolic line is $SU(1,1)/U(1) \simeq SL(2,\mathbbm{R})/SO(2)$.

One then chooses a certain representation $T: \mathfrak{u} \rightarrow 
\mathfrak{gl}(\tilde{\mathfrak{u}})$ and 
extends the Lie algebra $\mathfrak{u}$ to a solvable
Lie algebra $\mathfrak{l} = \mathfrak{u} \oplus \tilde{\mathfrak{u}}$,
with $[\tilde{\mathfrak{u}}, \tilde{\mathfrak{u}}] \subset \mathfrak{u}$.
The representation space $\tilde{\mathfrak{u}}$ is related to 
$\mathfrak{u}$ by an isomorphism of vector spaces $\gu \rightarrow
\tilde{\gu}$, $X\mapsto \tilde{X}$. The above basis is consistent with
this isomorphism, i.e.\ $G_0$ is mapped to $\tilde{G}_0$, etc.\  
The first complex structure $J_1$ on $\mathfrak{l}$ is then determined
by the condition that the restriction $\left. J_1\right|_{\mathfrak{u}}$ 
is the natural complex structure
on the K\"ahlerian Lie algebra $\gu$, together with  
the property that $J_1 \tilde{X} = - \widetilde{J_1 X}$ for all
$X\in \gu$. The second complex structure is defined by $J_2 X = 
\tilde{X}$ and $J_2 \tilde{X} = - X$ for all $X\in \gu$.
The representation $T$ is chosen as a $Q$-representation, which means that
it satisfies certain conditions which ensure that 
$(J_1, J_2, J_3=J_1J_2)$ is a quaternionic-K\"ahler structure on the solvable Lie algebra $\mathfrak{l}$.

One possible alternative approach to our work on para-quaternionic-K\"ahler structures would have been to adapt Alekseevsky's method
using bases analogous to the basis $(G_0, G_1, H_0, H_1, \tilde{G}_0,
\tilde{G}_1, \tilde{H}_0, \tilde{H}_1)$. However, the basis would
have needed to be adapted to the different scalar products, so that
we would have needed to work with three 
different bases, depending on the values
of $\epsilon_1$ and $\epsilon_2$. The advantage of the basis ${\cal T}$
is that it can be used in all three cases. Moreover, this basis 
is natural from the point of view of dimensional reduction in
supergravity.

\subsection{Conjugate Iwasawa subgroups vs disjoint open $L$-orbits}

For completeness we will explain the relation between, on the one
hand,  the open orbits
$M'=L' \cdot o$ of the canonical base point $o\in S = G/H$ under 
subgroups $L' = C_a(L) = a L a^{-1} \subset G$ ($a\in G$) conjugate
to the standard Iwasawa subgroup $L\subset G$ and, on the other hand,
the $L$-orbits $L\cdot o'$ of different points $o'\in S$. Notice that
the orbits $M'=L'\cdot o$ and $L\cdot o'$, $o'=a^{-1}o$, are related by
\[
M' = L_a(L\cdot o')  \;,
\]
where $L_a: S \rightarrow S$ is the diffeomorphism given by the 
$G$-action on $S$. 
\bp
Let $M'=L'\cdot o\;, M''=L'' \cdot o \subset S$ be the two open
orbits constructed in Sections \ref{IdTS} and \ref{IdST}. Then
the corresponding open $L$-orbits $L\cdot o'$ and $L\cdot o''$
are disjoint.  
\ep
\pf
Let us denote by $a',a'' \in G$ elements such that $L'= C_{a'}(L)$, 
$L'' = C_{a''}(L)$, $o'=(a')^{-1} \cdot o$ and $o'' = (a'')^{-1} \cdot o$. 
Recall that the open orbits $M'$ and $M''$ give rise
to left-invariant metrics $g'=(\phi')^* g_S$ and $g'' = (\phi'')^* g_S$ 
on $L$, where $\phi': L \rightarrow M'\;, \phi'(x)=C_{a'}(x) \cdot o$, and
$\phi'': L \rightarrow M''\;, \phi''(x) = C_{a''}(x) \cdot o$,  
for all $x\in L$.  
If $L\cdot o'$ and $L \cdot o''$ are not disjoint,
then they coincide, and $o'' \in L \cdot o'$. This means that there 
exists $a\in L$ such that $o'' = a o'$. Now we show that this implies
that the left-invariant metrics $g'$ and $g''$ on $L$ are related by
\be \label{conj}
g''=C_{a^{-1}}^* g' \;.
\ee
This follows from the equation 
\begin{equation}
\label{phiprimeprime}
\phi'' = L_{b} \circ \phi' \circ C_{a^{-1}} \;,\;\; b=a'' a (a')^{-1} \;, 
\end{equation}
which we will prove below.
In fact, using the $G$-invariance of $g_S$,  (\ref{phiprimeprime}) implies that 
\[
g'' = (\phi'')^* g_S = (C_{a^{-1}}^* \circ (\phi')^*)(L_b^* g_S) 
= (C_{a^{-1}}^* \circ (\phi')^*)(g_S) = C_{a^{-1}}^* g' \;.
\]
Now we prove (\ref{phiprimeprime}). We compute for $x\in L$:
\begin{eqnarray}
\phi'(x) &=& C_{a'}(x) \cdot o = ({a'}x)\cdot o'\,, \label{phiprimex} \\
\phi''(x) &=& C_{a''}(x) \cdot o = ({a''}x)\cdot o'' = ({a''}x a)\cdot
o' = ({a''} a C_{a^{-1}}(x)) \cdot o' \nonumber\\
&=& (a'' a (a')^{-1} a' C_{a^{-1}}(x)) \cdot o'
\stackrel{(\ref{phiprimex})}{=} b  \phi'(C_{a^{-1}}(x)) \;.
\end{eqnarray}
This proves (\ref{phiprimeprime}) and (\ref{conj}), under the assumption that the
orbits $L\cdot o'$ and $L\cdot o''$ 
are not disjoint. On the other hand, we know
from Proposition \ref{PropNonAut} that $g'$ and $g''$ are not
related by an inner automorphism of $L$. Therefore the orbits are
necessarily disjoint.  
\qed
\bc
The Iwasawa subgroup $L\subset G= G_{2(2)}$ acts with at least
two open orbits on $S=G_{2(2)}/(SL(2)\cdot SL(2))$.
\ec

\subsubsection*{Acknowledgements}

This work was
partly supported by the German Science Foundation (DFG) under the 
Collaborative Research
Center (SFB) 676 Particles, Strings and the Early Universe.\\
We would like to thank Owen Vaughan for useful discussions.
The work of T.M.\ is supported in part by STFC  
grant ST/G00062X/1.
The work of P.D.\ is supported by 
STFC studentship ST/1505805/1. 
T.M.\ thanks the Mathematics 
Department of the University of Hamburg for Hospitality and 
Support during various stages of this project. We thank Boris Pioline
for bringing reference \cite{Berkooz:2008rj} to our attention.

\providecommand{\href}[2]{#2}\begingroup\raggedright\endgroup

\end{document}